\documentclass[useAMS,usenatbib]{mn2e}

\usepackage{aas_macros,epsfig}

\title[Properties of Distant M-Stars]{M-Dwarfs at Large Heliocentric Distances}
\author[E R Stanway et al.]{Elizabeth R.~Stanway$^{1}$\thanks{email: E.R.Stanway@Bristol.ac.uk}, Malcolm N. Bremer$^{1}$, Matthew D. Lehnert$^{2}$, John J. Eldridge$^{3}$
\\$^1$H H Wills Physics Laboratory, Tyndall Avenue, Bristol, BS8 1TL, UK
\\$^2$Laboratoire d'Etudes des Galaxies, Etoiles, Physique et Instrumentation GEPI, Observatoire de Paris, Meudon, France
\\$^3$Astronomy Research Centre, School of Maths \& Physics, Queen's University Belfast, BT7 1NN, Northern Ireland, UK}

\begin{document}

\date{Accepted 2007 November 9.  Received 2007 November 9; in original form 2007 April 30}

\pagerange{\pageref{firstpage}--\pageref{lastpage}} \pubyear{}

\maketitle

\label{firstpage}

\begin{abstract}
  We present an analysis of the faint M star population seen as
  foreground contaminants in deep extragalactic surveys. We use
  space-based data to separate such stars from high redshift galaxies
  in a publically-available dataset, and consider the photometric
  properties of the resulting sample in the optical and infrared. The
  inferred distances place these stars well beyond the scale height of
  the thick disk. We find strong similarities between this faint
  sample (reaching $i'_{AB}=25$) and the brighter disk M dwarf
  population studied by other authors. The optical-infrared properties
  of the bulk of our sources spanning 6000\AA-4.5$\mu$m are consistent
  with those 5-10 magnitudes brighter.  We also present deep
  spectroscopy of faint M dwarf stars reaching continuum limits of
  $i'_{AB}\approx26$, and measure absorption line strengths in the
  CaH2 and TiO5 bands. Both photometrically and spectroscopically, our
  sources are consistent with metallicities as low as a tenth solar:
  metal-rich compared with halo stars at similar heliocentric
  distances. We comment on the possible MACHO identification of M
  stars at faint magnitudes.
\end{abstract}


\section{Introduction}
\label{sec:intro}

The advent in recent years of extremely deep imaging surveys, designed
to probe distant galaxies through multiwavelength imaging extending
deep into the infrared \citep[e.g. GOODS,][]{2004ApJ...600L..93G}, has
made possible studies of extremely faint, and often extremely red,
sources. In order to target high redshift galaxies, colour cuts have
been developed to exclude the majority of Galactic stars - including
the low metallicity dwarfs expected in the halo of our own galaxy.
Nonetheless, it has become clear that an unexpected number of
extremely red stars remain at faint magnitudes. The spectroscopic
component of high-redshift galaxy searches have confirmed the nature
of such stars, in several cases obtaining deep spectroscopy as well as
photometry \citep[e.g.][]{2004ApJ...607..704S,2005A&A...434...53V},
but the overall population has not been systematically characterised.

Low mass stars of class M or later comprise more than 80 per cent of the
stellar population of our galaxy. Faint stars of late-M, L and T
classes probe the tail of the stellar mass function, bridging the
divide between the main sequence and low mass brown dwarfs. Such stars
have unusual colours, arising from the presence of deep molecular
absorption bands in their cool atmospheres.

The faintness of these stars has made their investigation challenging,
particularly with increasing distance from our sun. Efforts to
determine the properties of such stars in the halo and thick disk have
concentrated on nearby stars with high proper motions indicative of a
halo origin \citep[e.g.][]{1999AJ....117..508G}. While deep surveys of
the M star population in order to identify distant examples have been
undertaken with WFPC2 on the {\em Hubble Space Telescope} ({\em HST}),
reaching $I_{AB}$=23.2 \citep{2001ApJ...555..393Z}, these have been
limited to wavebands shortwards of 9000\AA\ where the coolest stars
have very little flux.

Meanwhile, the more local population - late M stars at disk
metallicities within 1.5\,kpc of the sun - have been the subject of
detailed study in recent years. The Sloan Digital Sky Survey
\citep[SDSS,][]{2000AJ....120.1579Y} is relatively shallow in the
optical bands, reaching an $I$ band limit of $i'\approx20$. However,
the addition of a redder $z'$ band in the optical to similar depths has enabled the
photometric selection of cooler, class L and T stars as
sources with large flux decrements between adjacent passbands, caused
by sharply-defined, strong absorption and transmission features. The
$z'$-band imaging also allows the characterisation of late M stars by
their $i'-z'$ colour. Such surveys have enabled templates based on
thousands of local (mainly disk) dwarf stars to be constructed
\citep{2002AJ....123.3409H,2007AJ....133..531B}.

At the same time, advancements in infrared detector technology have
made it increasingly easy to study these faint, cool stars at
wavelengths longwards of the optical.  Their near and mid-infrared
bands are dominated by molecular absorption, leading to extreme
colours in infrared bands \citep{2006ApJ...651..502P}. Again, such
studies have been limited to local stars, in order to attain the
maximum possible signal to noise.

However, given the lower typical metallicity of stars in the halo when
compared to the galactic disc, the characteristics of local
disk-dominated M star samples may well be unrepresentative of fainter
sources at larger distances and the colour selection methods applied
locally might be expected to detect nothing at halo distances. 

In this paper we present an analysis of faint low mass stars, selected
for their extreme colours between two adjacent bands in high spatial
resolution, deep optical imaging. Such a colour selection is expected
to select either relatively-nearby late M stars at sub-solar metallicities
or near-solar metallicity early M stars at the large distances more
normally associated with the Galactic halo, if such a population
exists. These two alternatives cannot be distinguished on the basis of 
a single red colour, but can be explored using spectroscopy and photometry
probing well into the infrared.

In section \ref{sec:photometry} we present the data used in this
analysis and we discuss the selection of stellar candidates in section
\ref{sec:phot-sel}. In sections \ref{sec:halo-pop} and
\ref{sec:spitzer-properties} we discuss the optical and infrared
properties of our sample, and in section \ref{sec:cooler} we examine
the photometry of cooler stars in the same fields. In section
\ref{sec:spect} we discuss existing spectroscopy of such faint stars,
including a spectroscopic sample presented here for the first time,
and spectroscopic abundance indicators that may hint at a typical
metallicity for this population.  Finally, in section
\ref{sec:halo-pop} we discuss the interpretation of the overall
properties of the faint, cool star population including their
metallicity and distance distribution.

All magnitudes in this paper (optical and
infrared) are quoted in the AB system \citep{1983ApJ...266..713O}.


\section{Datasets used in this Analysis}
\label{sec:photometry}

\subsection{Filters and Photometric Systems}

We choose to use the AB system \citep{1983ApJ...266..713O} when
expressing magnitudes in this analysis. The reason for this is twofold.
Firstly the AB system is widely used by extragalactic astronomy and
the surveys that generate much of the deepest optical imaging,
including the data discussed below. The AB magnitude scheme is also
used in more local large spectroscopic samples such as the Sloan
Digital Sky Survey (SDSS).
Secondly, AB magnitudes are also constructed on a physical basis. A
source flat in flux as a function of frequency
(f$_\nu\propto\lambda^{-2}$) has zero colour in all bands.  



Colours are
calculated throughout this paper by convolving available observational
and theoretical spectra with the filter transmissions concerned
(including the instrument and detector response) to accurately account
for filter-specific effective wavelengths and full-width half-maxima.
Our main reference spectra for faint stars the observationally-derived
spectra of \citet{2007AJ....133..531B} which were constructed for each
subtype of class M and later stars from a total sample of over 700
observed spectra taken by the SDSS. As figure \ref{fig:templates}
illustrates, these template spectra trace the same stellar locus as
those of \citet{1998PASP..110..863P}, although slight deviations are
seen between the two template sets, particularly at late spectral
types.



\begin{figure}
\includegraphics[width=0.95\columnwidth]{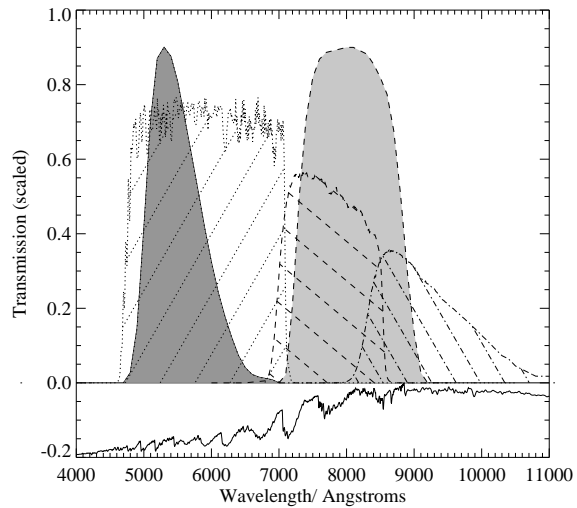}
\caption{The  transmission profiles of the  $F606W$
  ($v$, dotted), $F775W$ ($i'$, dashed) and $F850LP$ ($z'$, dot-dash) wavebands, and of the
  \citet{1990PASP..102.1181B} $V$ (dark grey) and $I$ (light grey)
  filters. Offset and plotted below is the spectrum of an M4 star.}\label{fig:filter_profiles}
\end{figure}

Use of the AB (rather than Vega) magnitude system introduces a
blueward shift of 0.43 magnitudes in the ``Standard'' Bessell $V-I$
colour, which accounts for the spectral slope of $\alpha$ Lyr at long
wavelengths.  In addition the $F606W$ ($v$) and $F775W$ ($i'$) filter
set has redder central wavelengths than their Bessell filter
counterparts, and the $F606W$ filter is broader than the ``Standard''
$V$ as figure \ref{fig:filter_profiles} illustrates. 



\subsection{Surveys}

The analysis presented in this paper utilises two deep,
multi-wavelength imaging surveys in order to combine their different
strengths.

The photometric selection of cool dwarf stars parallels that of high
redshift galaxies in many respects.  In both cases the accurate
characterisation of the sources, and their reliable separation from
other sources with similar optical colours, requires multiwavelength
imaging spanning a long baseline from 5000\AA\ to 3.6$\mu$m, and high
angular resolution impossible to obtain from the ground without
the use of adaptive optics.

The Great Observatories Origins Deep Survey\footnote{http://www.stsci.edu/science/goods/} (GOODS)
satisfies the requirements for such a selection. In addition to an
abundance of auxiliary data, the survey comprises two primary data
sets, optical and infrared, each of which surveys a field in the
northern hemisphere (GOODS-N) and a second in the south (GOODS-S). All
GOODS data is publically released to the astronomical community in the
form of fully reduced images.

The optical (0.3-1$\mu$m) component of GOODS \citep{2004ApJ...600L..93G} consists of
deep optical imaging in the $F435W$ ($b$), $F606W$ ($v$), $F775W$
($i'$) and $F850LP$ ($z'$) bands, each reaching depths $>28$th
magnitude as shown in table \ref{tab:depths}, obtained using the wide
field channel of the Advanced Camera for Surveys (ACS) on the {\em
  Hubble Space Telescope} (HST). The imaging has been drizzled to a
pixel scale of 0$.''$03 per pixel, with a point source FWHM of
0$.''$05 in the $i'$ band. In addition, the GOODS team have made
publically available catalogues of the source photometry and properties
in each field, generated using the SExtractor software package
\citep{1996A&AS..117..393B}, and with detection parameters tuned to
the survey depth and resolution. In this analysis we utilise the
Kron-radius based automatic magnitudes reported by SExtractor (i.e.
MAG\_AUTO). These magnitudes follow the curve of growth of the source
profile, and are equivalent to a corrected aperture magnitude for an
unresolved source. We use version r1.1 of the GOODS
catalogues\footnote{http://archive.stsci.edu/prepds/goods/}.

The second primary dataset generated by the Great Observatories
Origins Deep Survey comprises deep {\em Spitzer Space Telescope}
infrared (3-10$\mu$m) imaging of the fields targeted by {\em Hubble}
\citep{2005AAS...207.9507D}.  The IRAC instrument was used to survey
the GOODS fields at 3.6, 4.5, 5.8 and 8.0\,$\mu$m, with a pixel scale
of 1.2$''$/pix to a depth $>$26th magnitude in the first two
bands and $>$24th magnitude at longer wavelengths as shown in table
\ref{tab:depths}\footnote{The GOODS programme also obtained imaging
  with the MIPS instrument that is not discussed here}. As discussed in
section \ref{sec:spitzer-properties}, confusion is a significant issue
for sources at the faint magnitudes of our targets, and in imaging of
this depth. As a result, the 12 arcsecond (10 pixel) apertures
recommended for IRAC photometry in sparse fields seldom escape
confusion in the GOODS images. \citet{2007astro.ph..1725V} determined
that smaller apertures with a diameter of 4.5 arcseconds were suitable
for compact sources in the GOODS images, and could be reliably
corrected for flux in the wings of the point spread function. We apply
the same prescription for aperture magnitudes as
\citeauthor{2007astro.ph..1725V}, correcting to total magnitudes using
the offsets given in table \ref{tab:depths}.

\begin{table*}
\begin{tabular}{lcccccccc}
Band  & b & v & i' & z' & 3.6\,$\mu$m & 4.5\,$\mu$m & 5.8\,$\mu$m & 8.0\,$\mu$m\\
\hline\hline
Depth & 28.87 & 29.46 & 28.85 & 28.55 & 26.96 & 26.38 & 24.50 & 24.37\\
$m_{tot}-m_{ap}$ & - & - & - & - & -0.180 & -0.180 & -0.326 & -0.418\\
\end{tabular}
\caption{The 3\,$\sigma$ depth of the GOODS imaging used in this analysis,
  and aperture corrections applied in the Spitzer wavebands. The limits
  for optical bands are given in a 1 arcsecond aperture and those for the
   Spitzer bands in a 4.5 arcsecond aperture, corrected to total 
   magnitude.}\label{tab:depths}
\end{table*}

Both GOODS fields look out of the plane of the galactic disc and away
from the galactic centre, with galactic coordinates of $l$= 125.865,
$b$=+54.808 for the GOODS-N field and $l$=223.57, $b$=-54.43 for the
GOODS-S. Both fields were also selected to lie in regions of low
Galactic extinction, selecting against structures in the galactic
disk.

The spectroscopic analysis of faint M stars presented in section
\ref{sec:spect} is also based in part on GOODS data. Several of
the stars identified in section \ref{sec:phot-sel} have deep optical
spectroscopy taken with FORS2 at the Very Large Telescope (VLT) as
part of the ESO GOODS survey \citep{2005A&A...434...53V}.

In order to improve the statistical significance of our analysis, we
supplement the ESO GOODS data with further M class star spectra
observed as part of the BDF project. The BDF survey comprises deep
multicolour imaging and spectroscopic follow-up of red sources in four
near-contiguous fields, applying an $R-I$ colour selection to identify
galaxies at $z>5$, and reaching a limiting depth of $I_{AB}=26.3$. The
first of these fields, BDF1, was described in
\citet{2003ApJ...593..630L} while the remaining three fields, BDF2-4,
were observed with an identical strategy and will be described in a
forthcoming paper. These data are not used in the photometric analysis
since with ground-based seeing it is impossible to separate mid-M
stars from high redshift galaxies on the basis of optical colours
alone (without $B$ band imaging reaching some four magnitudes deeper
than the $I$ band limit). However, the high spectroscopic completeness
of this survey for objects with extreme $R-I$ colours (essentially the
same selection function as the {\em HST}/ACS $V-I$ selection discussed
in section \ref{sec:phot-sel}) makes it extremely useful for the
\textit{a posteriori} analysis of stellar spectra.

The FORS2 $R$ (R\_SPECIAL, ESO number 76) and $I$ (I\_BESS, ESO number
77) band filters used for imaging in the BDF fields were selected to
be sharp sided and to have minimal overlap. Despite the different
naming, the wavelength coverage of the $R$ band at 6550\AA\ 
(FWHM=1650\AA) overlaps significantly with that of the $v$ band at
5907\AA\ (FWHM=2343\AA). Hence a colour cut of $(R-I)_{AB}>1.5$
reproduces the {\em HST}/ACS $v-i'>1.3$ selection function for stars
discussed in section \ref{sec:phot-sel}, selecting class M3-M4 and
later.

\section{A Photometric Sample of Faint M-dwarfs}
\label{sec:phot-sel}

Class M, L and T stars have an underlying red continuum due to
their cool temperatures. Their spectrum in the yellow and red is 
dominated by jagged molecular absorption bands that further redden the 
$v-i'$ and $V-I$ colours.
As a result, such cool stars can have extreme colours in
adjacent filters and are regularly selected by `dropout' surveys
searching for high redshift sources (which show a similar drop between
bands due to neutral hydrogen absorption in the intergalactic medium).
We note that although `dropout' is a historical term used by
convention in the selection of high redshift galaxies, neither high
redshift galaxies nor cool stars are expected to have zero flux in the
bluewards band. As a result, whether the sources show a finite colour
or drop entirely below the detection limit in the bluewards band
depends on the relative depth of the imaging.  Hence the term `break'
(as in Lyman break galaxy) or `drop' is more technically correct.
Sources are required to satisfy the criterion of a significant
spectral break, measured through their photometry, but not to remain
undetected in any given band. Nonetheless, for historical reasons, we
continue to use the term `dropout' while cautioning readers that its
literal meaning is inaccurate.

\begin{figure}
\includegraphics[width=0.95\columnwidth]{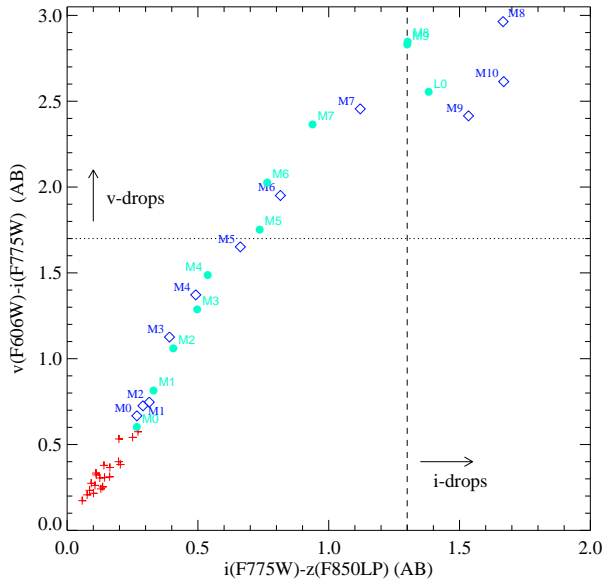}
\caption{The $v-i'$ and $i'-z'$ colours of cool (class M and L) stars,
  calculated from empirical template spectra and convolved with the
  {\em HST}/ACS filter set. Solid points indicate the cool star sample
  of \citet{2007AJ....133..531B}, diamonds indicate the M star spectra
  of \citet{1998PASP..110..863P} while crosses are K stars from the
  same source. M class stars increase linearly in $i'-z'$ colour with
  increasing subclass, although we note that the two sets of empirical
  spectra differ significantly at certain
  subclasses.}\label{fig:templates}
\end{figure}

In order to select mid-M class stars and later, we apply the same
$v$-drop selection function commonly used to select high redshift
galaxies. As figure \ref{fig:templates} illustrates, a colour cut of
$(v-i')_{AB}>1.7$ \footnote{which has been used to identify the
  Lyman-$\alpha$ spectral break in galaxies at $4.8<z<5.8$}
effectively selects stars of class M3 and later at solar metallicity
(see section \ref{sec:halo-pop} for discussion for extreme low
metallicity populations which will fall out of our colour selection).
We note that we are likely to be insensitive to later spectral types:
despite their red colours in $v-i'$, these sources are generally too
faint in $i'$ to be detected to our survey limit in the $i'$-band.
Beyond M6, the almost-linear increase in $v-i'$ colour begins to
plateau, reducing the ability of this colour to distinguish between M
class subtypes. Adding a second, redwards colour such as $i'-z'$
improves the discrimination between cool stars. Hence in order to
examine the coolest stars in our survey fields we also consider a
$z'$-band selected sample of stars satisfying $(i'-z')_{AB}>1.3$
\footnote{which identifies galaxies at $5.6<z<6.5$ or class L and T
  stars}.

\begin{figure}
\includegraphics[width=0.95\columnwidth]{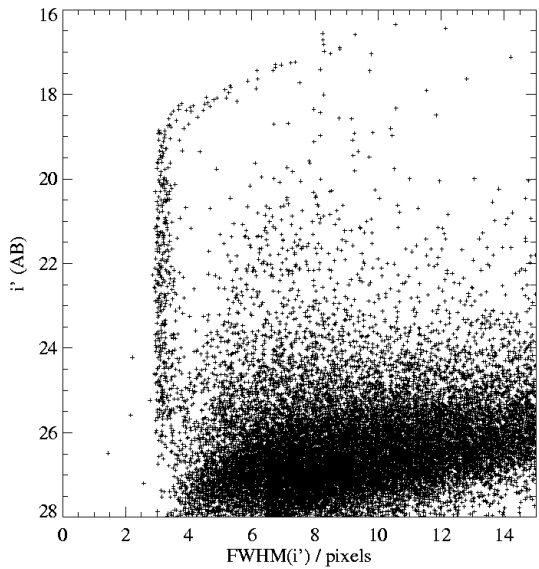}
\caption{The distribution of Gaussian FWHM determined for sources
  in the GOODS fields. The stellar locus is clearly separated from the
  distribution of extended galaxies for $i'_{AB}<25$ with very few
  ambiguous sources. We apply a criterion of FWHM $<4$\,pixels to
  select stars. Note that sources saturate in the GOODS imaging at
  19th magnitude}\label{fig:fwhm_mag}
\end{figure}

The resulting sample is expected to contain several distinct
populations: the targeted stars, high redshift ($z>5$) galaxies and
quasars which show a strong break at $\lambda_{rest}=$1216\AA, and old
elliptical galaxies at $1<z<2$ which break at
$\lambda_{rest}=$4000\AA.  The majority of contaminant galaxies can be
identified by their extended half light radii in the {\em HST}/ACS
data. All confirmed $z>5$ galaxies selected by such a colour cut
criterion are resolved in space-based imaging, albeit barely in
several cases
\citep{2003MNRAS.342L..47B,2004MNRAS.347L...7B,2004ApJ...607..704S,2004ApJ...604L..13S,2004ApJ...611L...1B}.
Similarly, galaxies at $1<z<2$ are typically well resolved by HST. As
figure \ref{fig:fwhm_mag} illustrates, unresolved sources separate
cleanly from extended sources in the GOODS imaging for sources
brighter than $i'_{AB}=25$ (or for $z'_{AB}=26$ in the case of
$z'$-selected sources).

Clearly high redshift quasars cannot be distinguished from stars by
their half light radii, since both populations are expected to be
unresolved. The predicted number of quasars at the faint magnitudes
probed here is expected to be small. \citet{2007astro.ph..1724D}
identified only one active galactic nucleus (AGN) in a highly
spectroscopically complete survey of dropout objects examining 450
arcmin$^2$ to a depth of 26th magnitude in $I$\footnote{Similar to the
  surface densities for faint AGN found by \citet{2005ApJ...634L...9M}
  and \citet{2007astro.ph..1515S}}. That source was slightly resolved
in {\em HST}/ACS data, with comparable contributions to the rest-frame
ultraviolet flux from the AGN and a starburst in the host galaxy.
Hence we might expect to detect one AGN in the 300\,arcmin$^2$ GOODS
survey area, either unresolved or marginally resolved. Fortunately,
such a source should be distinguishable based on its infrared colours
\citep{2006astro.ph..8603S} as discussed in section
\ref{sec:spitzer-properties}.

In order to select low mass stars in the available ACS imaging with
minimal contamination from extragalactic sources, we apply an
additional selection criterion, requiring that: FWHM$_{i'} \le 4.0$
pixels (0$''.$12), and we truncate our $v$-drop selection at
$i'_{AB}=25$ and our $i'$-drop selection at $z'_{AB}=26$.

Each source was visually inspected to ensure that the photometry was
not contaminated by neighbouring objects.  In this deep imaging, the
area of sky affected by sources above the detection limit is $2$\%.

The resultant number counts of dropout-selected stars is shown in
table \ref{tab:number}. In this analysis we focus on the M stars in
the sample since these are both numerous, allowing a statistical
analysis, and relatively bright, allowing spectroscopic follow-up.

\begin{figure}
\includegraphics[width=0.95\columnwidth]{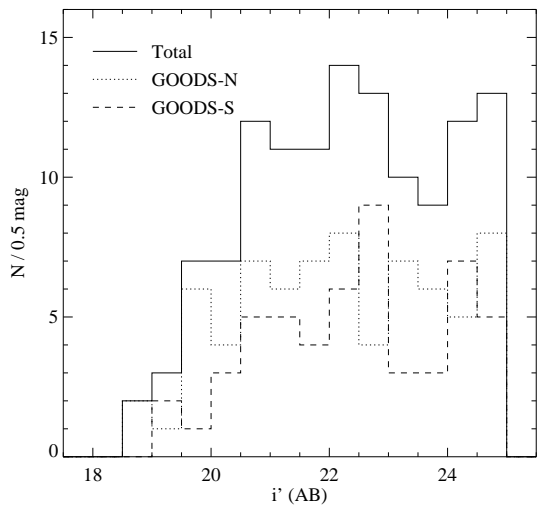}
\caption{The apparent magnitude distribution of $v$-drop stars
  in the GOODS field. The decline in number counts at bright
  magnitudes arises both due to the small volumes probed at bright
  magnitudes and due to the potential saturation of brighter sources
  in the deep GOODS imaging. There is no obvious decline in the number
  of sources seen at faint magnitudes. Each field surveys an area of
  150\, arcmin$^2$.}\label{fig:number}
\end{figure}

 While the $i'$-drop selected sample is too
small to derive reliable statistics, the magnitude distribution of the
$v$-drop sample (i.e. M stars) is illustrated in figure
\ref{fig:number}.  There is no clear trend in the number density of
cool stars with apparent magnitude. The northern GOODS field is 34\%
more abundant in low mass stars than the southern field, despite their
similar areas and orientation with respect to the galactic disk.

\begin{table}
\begin{tabular}{lccc}
Field & Area/sq arcmin & $N(v$-drop)& N($i$-drop)\\
\hline\hline
GOODS-S & 150 & 53 & 7 \\
GOODS-N & 150 & 71 & 5 \\
\end{tabular}
\caption{The number of unresolved sources in the deep
  {\em HST}/ACS imaging satisfying our colour selection 
  criteria as described in section \ref{sec:phot-sel}.}\label{tab:number}
\end{table}

\section{The Optical Properties of Faint M-dwarfs}
\label{sec:halo-pop}

The selection of stars by photometric methods relies on a combination
of the red optical spectrum that arises from a cool blackbody with the
abrupt absorption edges of metal molecules. In the absence of
significant metal enrichment, the colours of stars at a given
temperature are likely to be less extreme since the strength of
molecular features is reduced.  Hence a sample of faint stars based on
extreme colours in two adjacent bands is expected to select either
relatively-nearby late M stars at low metallicities as well as
near-solar metallicity early M stars at the large distances more
normally associated with the Galactic halo, if such a population
exists.

The \citet{1995ApJ...445..433A} `NexGen' models and the Kurucz ATLAS9
models \citep{2004astro.ph..5087C,1996ASPC..108....2K} remain the most
recent stellar atmosphere models to attempt the temperature and
metallicity regime of interest here (although the ATLAS9 models in
fact do not survey T$<$3500K late M stars and do not account for the
important CaH bands in cool stellar spectra). As the recent analysis
by \citet{2004AJ....128..829B} shows, neither provides a satisfactory
fit to the spectra of cool stars. While the MARCS spectra produce a
somewhat better fit, the publically available library only extend as
cool as 4000K. In practise, no stellar models can reliably reproduce
the colours of cool stars.

On the other hand, there exist a number of published 'extreme' cool
stars known to be at significantly sub-solar metallicity
\citep[e.g.][]{1997AJ....113..806G,2007ApJ...657..494B}. Unfortunately
the photometry and/or spectral coverage of these observations
inevitably covers too short a wavelength range to reliably calculate
$v$, $i'$ and $z'$ colours simultaneously from the empirical spectra. 

As a result, while we tentatively consider the results of available
model templates as well as empirical spectra, we caution the reader
that such results should not be considered reliable. We note that
future interpretations of this population would benefit from improved
modelling and interpretation of these faint stars in the red spectral
region ($>$8000\AA) accessible to red-sensitive spectrographs.

We have determined the optical colours
expected for cool stars, in the same filter set employed by the GOODS
survey, over a range of metallicities. We use the M-star spectral
models of \citet{1995ApJ...445..433A}\footnote{In their `NextGen' form
  as implemented by \citet{1998A&AS..130...65L}} over the range
[Fe/H]=0. to -4.0 (where such models are available) and at four
different temperatures: T=2000K, 2500K, 3000K and 3500K.

\begin{figure*}
\includegraphics[width=0.95\columnwidth]{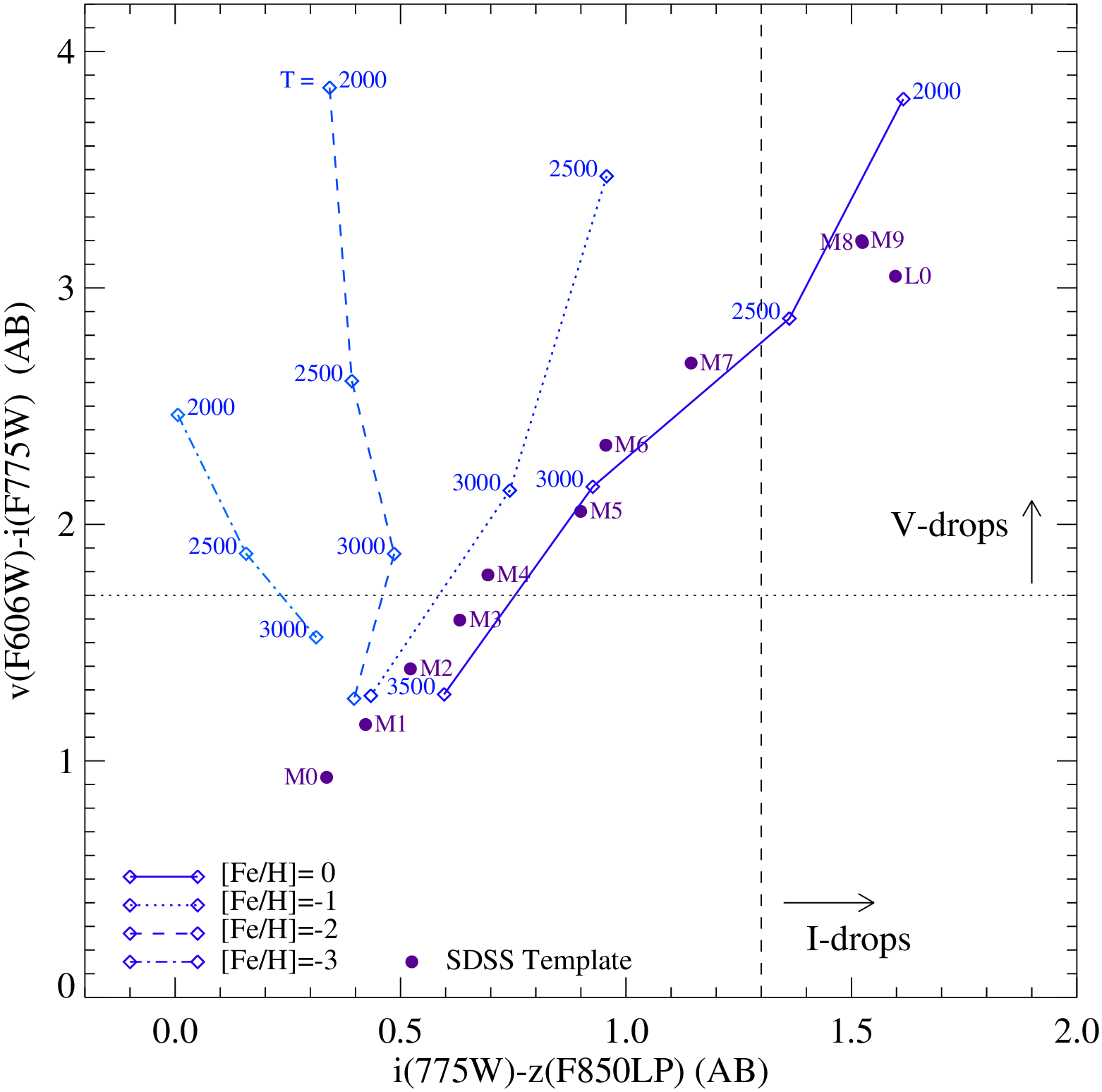}
\includegraphics[width=0.95\columnwidth]{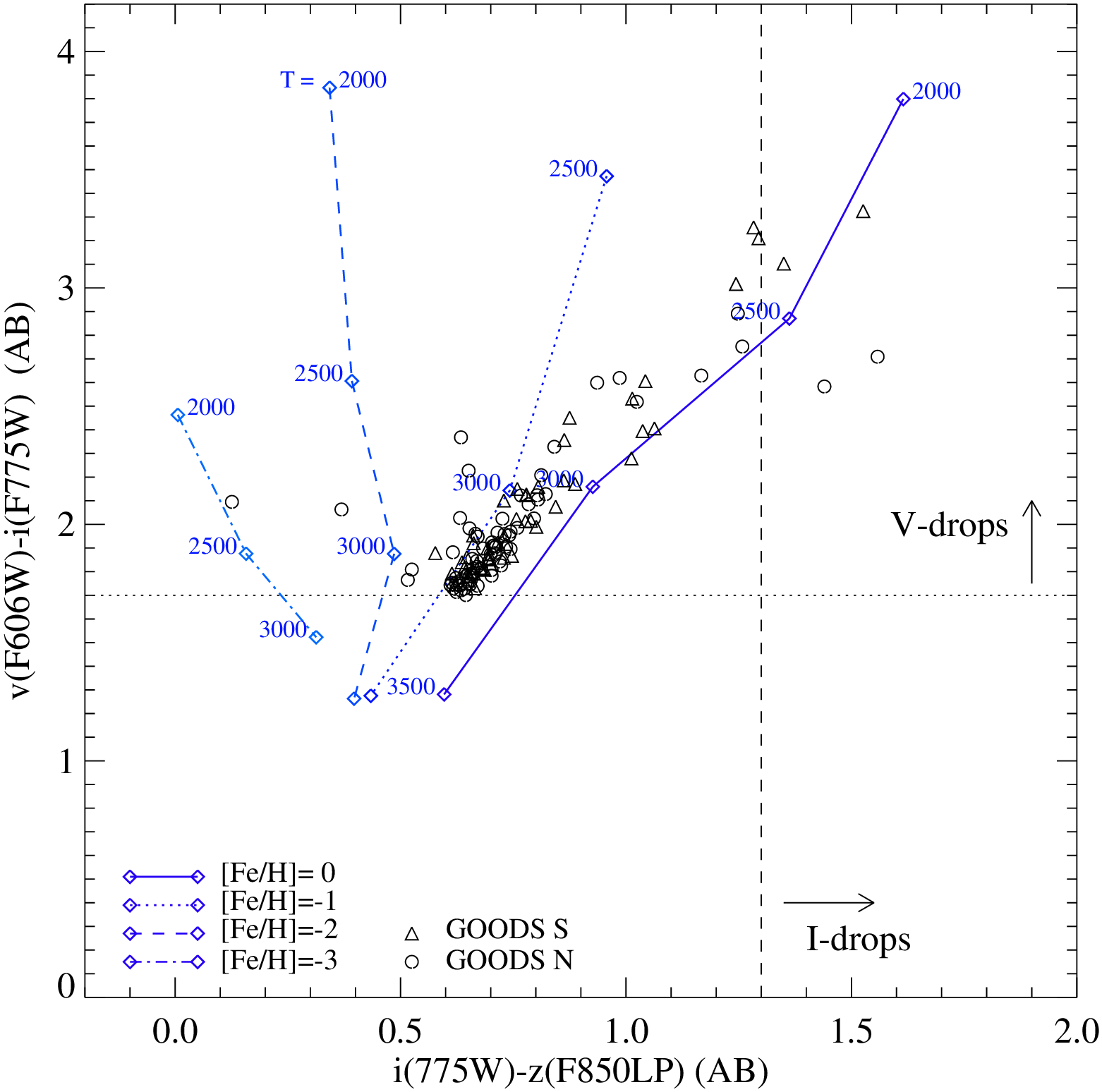}
\caption{The $v-i'$ and $i'-z'$ colours of model dwarf star
  atmospheres \citep[from][, triangles]{1995ApJ...445..433A},
  convolved with the {\em HST}/ACS filter set. Points at different
  temperature for the same metallicity are joined by lines and the
  temperature in kelvin is labelled. Figure a (left) overlays these
  tracks with the colours of moderate metallicity dwarf star templates
  derived by \citet{2007AJ....133..531B}, while figure b (right)
  overlays the colours of faint stars identified in section
  \ref{sec:phot-sel}. Photometric errors in both colours are
  $<$0.05\,mag.}\label{fig:metallicity}
\end{figure*}

Figure \ref{fig:metallicity} illustrates the colours predicted by the
`NextGen' models, and compares these with the colours of the SDSS
dwarf star templates derived by \citet{2007AJ....133..531B} which are
dominated by nearby disk stars and the photometric sample derived in
section~\ref{sec:phot-sel}. It is clear that while the $v-i'$ colour
fails to distinguish between different metallicity populations, the
addition of an $i'-z'$ colour may provide a crude measure of metallicity.
We note that a slight offset (approx 0.1 magnitudes in $i'-z'$)
appears to be required to match the model spectra to the observed
near-solar, old disk stellar sample at bright magnitudes.

Given the uncertainties in current models of M star atmospheres, we
compare their predictions with observational data on Galactic globular
clusters.  Observational evidence for the $i'-z'$ (or more generally
$I-Z$) colours of cool stars as a function of metallicity seems to be
poor.  However the $V-I$ colours of such stars (observed in the
Johnson-Cousins or Bessell systems and in Vega magnitudes) have been reported from observations of several
globular clusters. While few observations probe deep enough to
identify the base of the main sequence (as opposed to the main
sequence turn-off), some observations of low metallicity globular
clusters have characterised the main sequence ridge to very faint
magnitudes.

NGC 6397 is one such cluster, with a metallicity of [Fe/H]=-2.0 and
lying at a distance of 2.2\,kpc \citep{1996ApJ...468..655C}. The main
sequence in this source has been observed with {\em HST}/ACS down to a
limiting magnitude of $I_{F814W}\sim30$ (Vega) and is seen extending
redwards to $V_{F606W}-I_{F814W}\sim4$, the theoretical
hydrogen-burning limit at this metallicity
\citep{2006Sci...313..936R}.  At our colour cut of $v-i'$=1.7
(approximately $V-I_\mathrm{Vega}=2.6$), the main sequence in this
cluster corresponds to an observed magnitude of approximately
$I\_{814}=24.5$ (Vega), or an absolute magnitude limit of 12.4 in the
I band, while our detection limit of $i'_{AB}=25$ would only select
sources with $V-I_\mathrm{Vega}<3.0$.  This cutoff in apparent
magnitude implies that we are only sensitive to approximately class
M4-M5 stars at the distance and metallicity of NGC 6397
\citep{2002AJ....123.3409H} and would not expect to select any stars
with halo metallicities at greater distances.  This emphasises that
our colour cut allows only for a narrow range of low metallicity
objects which are sufficiently red and sufficiently bright to meet our
colour selection criteria.



\begin{figure}
\includegraphics[width=0.95\columnwidth]{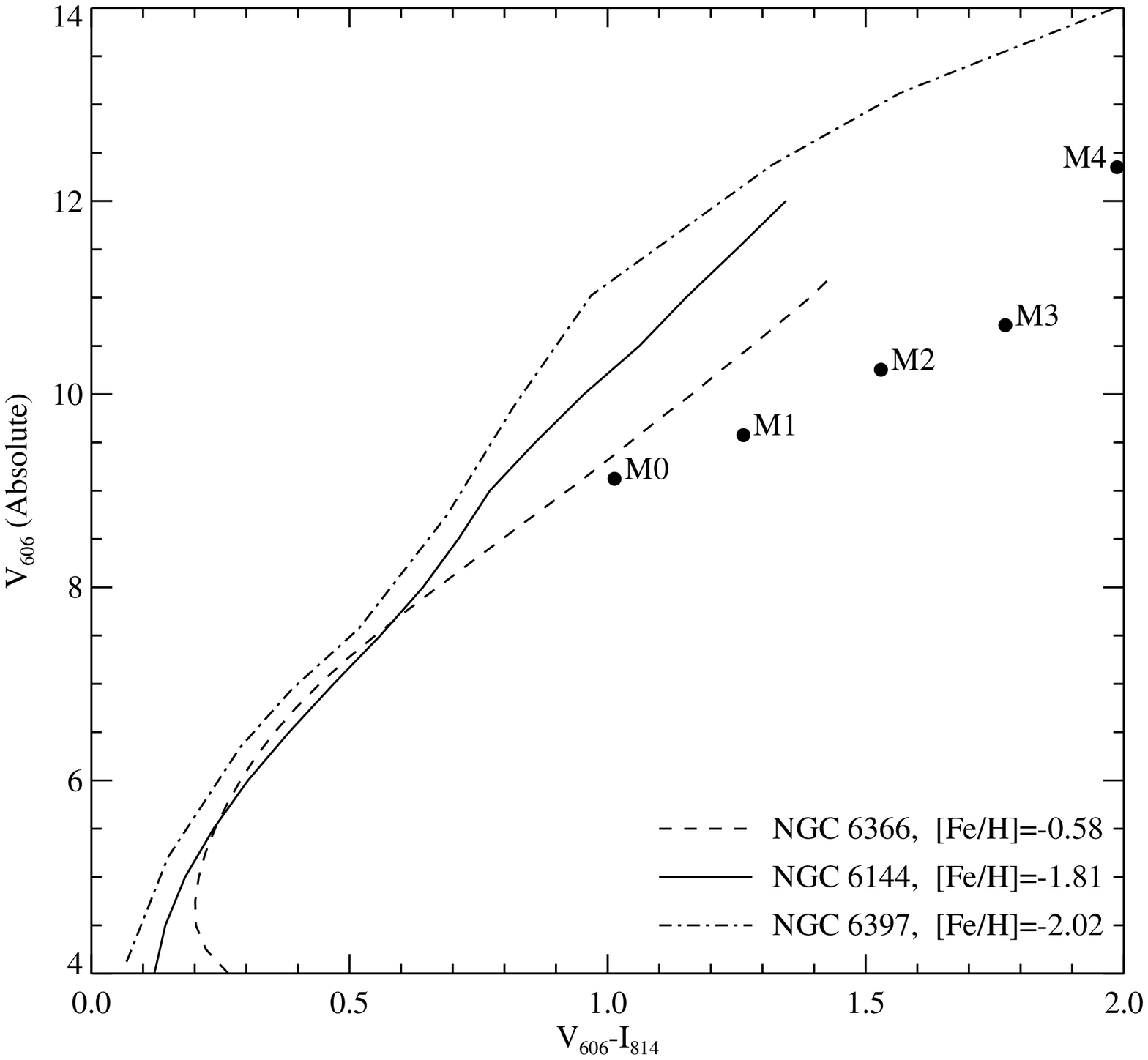}
\caption{The colour-absolute magnitude sequence of three globular clusters
  \citep{2006Sci...313..936R,2007AJ....133.1658S} and of the cool
  stars identified by the SDSS \citep{2007AJ....133..531B}. In each
  case information in the given references is used to correct sources
  to absolute magnitudes on the AB system and, where necessary, to
  correct for Galactic reddening. For ease of comparison with the
  globular cluster work, the SDSS stars are shifted into the HST ACS
  $F606W$ and $F814W$ bands. The data in NGC 6144 and NGC 6366 become
unreliable redwards of approximately $V-I=1.3$}\label{fig:globulars}
\end{figure}

In figure \ref{fig:globulars} we illustrate the effects of decreasing
metallicity on the main sequence colours of globular clusters with
published photometry. As discussed above, NGC 6397 is a distant
globular cluster with the low metallicity expected of halo stars. By
contrast NGC 6366 and NGC 6144 are two clusters with [Fe/H]=-0.58 and
[Fe/H]=-1.7 respectively, and published main sequence star photometry
to faint magnitudes observed with HST/ACS \citep{2007AJ....133.1658S}
as part of an ACS survey of globular clusters. While these data are
amongst the deepest available, they do not reach the red colours, and
hence faint $V$-band magnitudes of the M stars of interest here.
Nonetheless, it is interesting to note that the well-defined main
sequence in these globular clusters marks a gradual shift to bluer
$V-I$ colours with decreasing metallicity, and all three are bluer
than the old-disk stars measured by the SDSS.

The spectra of several very low metallicity extreme subdwarfs in the
field have now been published. \citet{2007ApJ...657..494B} demonstrate
the effects of decreasing metallicity on the optical spectra of late M
dwarfs. The weakening of metal oxide bands at 7200\AA, 7900\AA\ and
8500\AA\ with decreasing metallicity boosts stellar flux in the $i'$
and $z'$ bands.  In particular, the deep 7900\AA\ absorption feature
in the spectra of cool stars at disk metallicities suppresses the
$i'$-band magnitudes of these sources. Decreasing metallicity removes
this suppression, boosting the $v-i'$ colour of these sources while
decreasing their colour in $i'-z'$. While none of the published
extreme subdwarfs have published $z'$-band photometry, synthetic
colours calculated from the \citet{2007ApJ...657..494B} spectra
indicate that at these low metallicities, late M dwarfs are
approximately 0.2\,mag bluer in $i'-z'$ than typical old-disk M stars
(Bessell, private communication).

It is clear from the combination of
theoretical and observed data that most stars earlier than M3-M4 will
be omitted at low metallicities due to their blue colours, while later
M class stars are only likely to be detected relatively nearby due to
their faint magnitudes. Thus our selection preferentially identifies
stars with disk metallicities, while probing to halo distances.
This analysis of our colour selection function is supported by the
distribution of optical colours in our faint star sample as shown in
figure \ref{fig:metallicity}b. The majority of stars have colours consistent with
metallicities in the range {$-1<\mathrm{[Fe/H]}<0$} (i.e. similar to that of
the Galactic disk).

Two stars in the GOODS-N field show anomalously low $i'-z'$ colours
($i'-z'<0.  5$, $v-i'=2.1$) as might be expected at [Fe/H]$>$-2, and
both of these are relatively faint ($i'>23$). Comparison with the
colours of the extreme subdwarfs of \citet{2007ApJ...657..494B}
suggests that these might be identified as very low metallicity
sources, although both are somewhat redder in $v-i'$ than might be
expected. Either spectroscopy of these two blue stars, or optical
photometry of known extreme subdwarfs will be required to make a more
quantitative comparison.

\section{The Infrared Properties of Faint M-dwarfs}
\label{sec:spitzer-properties}

\begin{figure*}
\includegraphics[height=0.6\columnwidth]{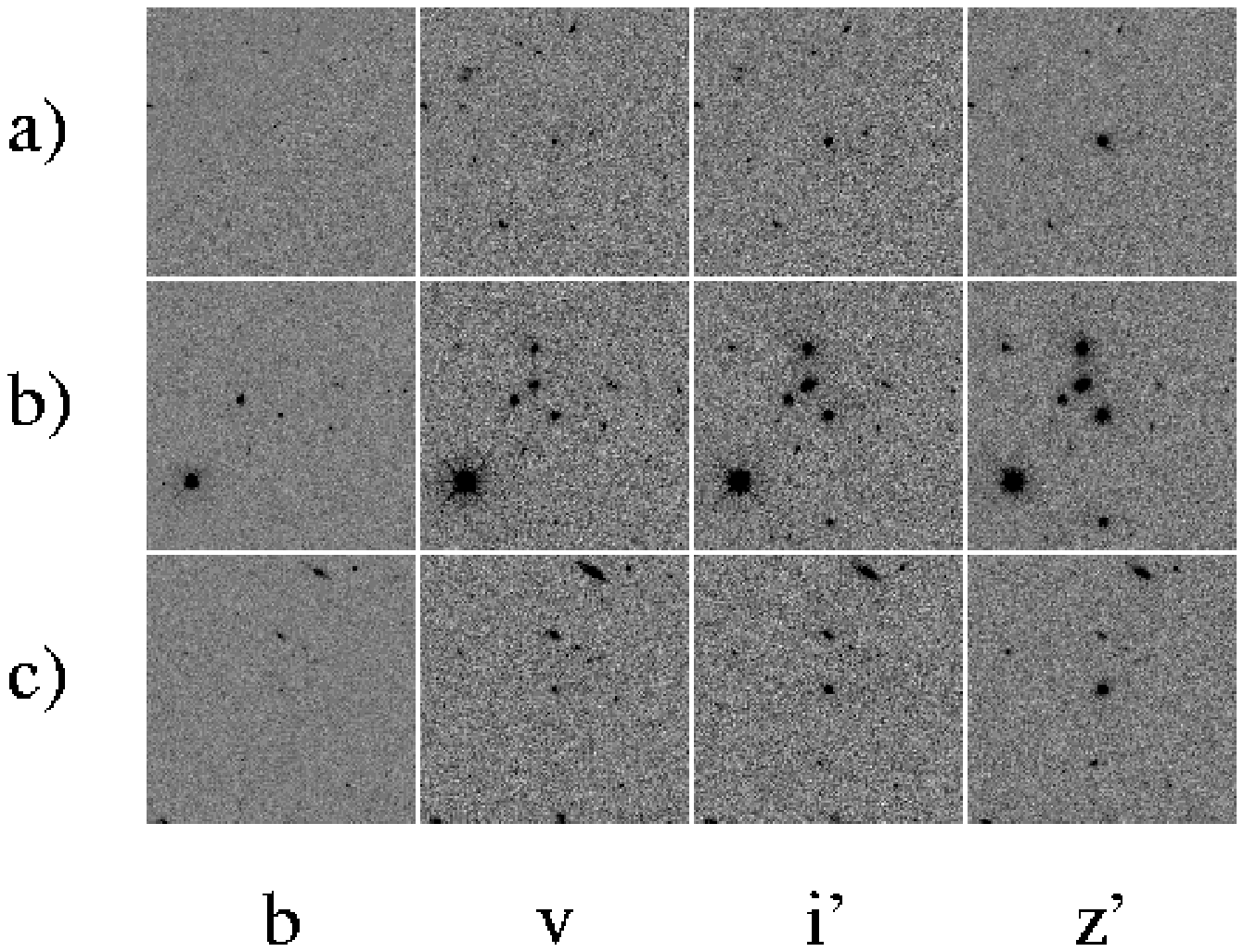}
\includegraphics[height=0.6\columnwidth]{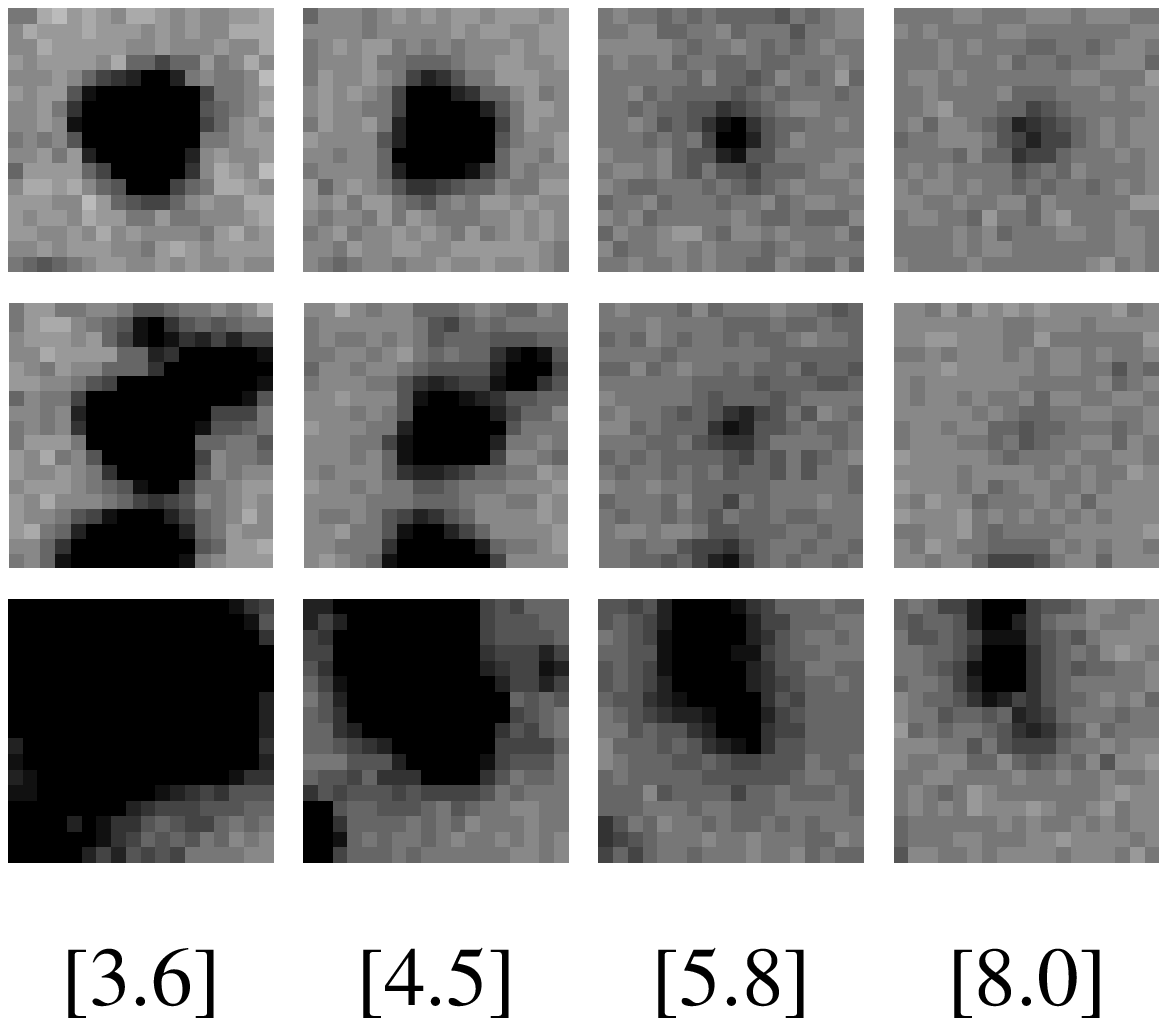}
\caption{Examples of GOODS/ACS selected sources in the
  GOODS/IRAC imaging. Some sources are isolated or have only much
  fainter neighbours (e.g. case a), while others are hopelessly
  confused in the IRAC bands (e.g. case c) with neighbours of
  comparable magnitude or brighter. Some sources are blended but not
  beyond hope of deconvolution (e.g case b). Boxes are 20$''$ to a
  side.}\label{fig:spitzer_examples}
\end{figure*}

Having identified the stars in HST data, we now consider the colours
of these unresolved sources in the infrared with {\em Spitzer} GOODS
data. The deep images are close to the confusion limit in the IRAC
bands, and many of the $v$-drop sources are confused in the imaging.
Of a total of 124 $v$-drop stars in the GOODS fields we classify 48
(39\%, figure \ref{fig:spitzer_examples} case a) as isolated, a
further 36 (29\%, figure \ref{fig:spitzer_examples} case b) as
confused but potentially deblendable and the remaining 40 (32\%,
figure \ref{fig:spitzer_examples} case c) as hopelessly confused. In
this analysis we use only isolated objects since the noise in
deconvolved magnitudes at this level is often dominated by residuals
from neighbouring objects rather than the target source.

\begin{figure}
\includegraphics[height=0.9\columnwidth]{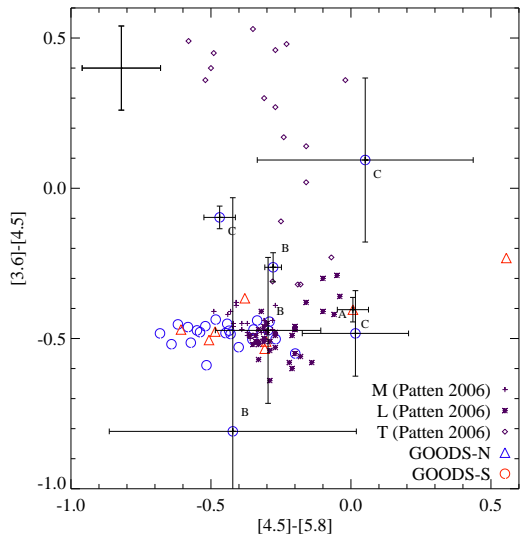}
\caption{The [4.5]-[5.8] versus [3.6]-[4.5] colours of isolated stars
  selected as $v$-drops in the IRAC GOODS fields. The cool star sample
  of \citet{2006ApJ...651..502P} which comprises M5-T9 stars at bright
  magnitudes is shown for reference. A typical uncertainty in the
  colours (dominated by uncertainty in the aperture correction and
  residual flux from nearby sources) is shown in the upper
  left.}\label{fig:spitzer_23v12}
\end{figure}

\begin{figure}
\includegraphics[height=0.9\columnwidth]{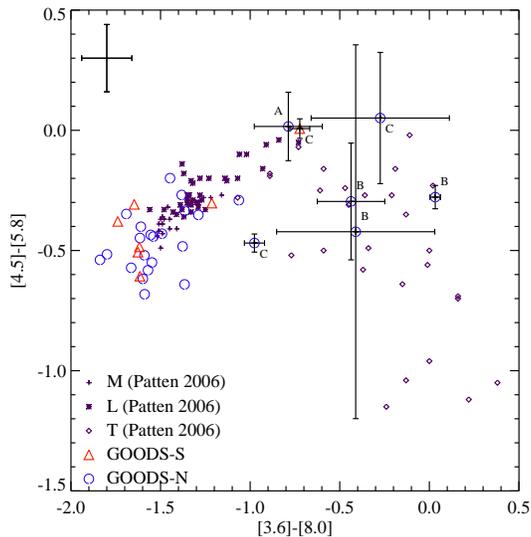}
\caption{The [3.6]-[8.0] versus [4.5]-[5.8] colours of isolated
  stars selected as $v$-drops in the IRAC GOODS fields. As in figure
  \ref{fig:spitzer_23v12}. Error bars are shown separately for 
  stars with anomalous colours in these bands (i.e. [3.6]-[8.0]$>$-1).
  These fall into three categories; A, B and C as described in section
  \ref{sec:spitzer-properties}.}\label{fig:spitzer_14v23}
\end{figure}

The infrared colours of the remaining sample are shown in figures
\ref{fig:spitzer_23v12} and \ref{fig:spitzer_14v23}, with the colours
measured for M, L and T dwarfs in the same bands by
\citet{2006ApJ...651..502P} shown for comparison. Working some three
magnitudes above the 3\,$\sigma$ background limit, the errors in the
colours are dominated by uncertainty in the aperture correction and
are set to an indicative 0.1 mag in each band.

The colours of cool stars in the infrared are dominated by the
relative strengths of molecular absorption bands in their atmospheres,
in particular those of CH$_4$ in the 3.6\,$\mu$m band, CO in the
4.5\,$\mu$m band and H$_2$O in the 5.8\,$\mu$m band.  

The infrared colours of our faint $v$-drop selected sources are very
similar to those of the brighter population studied by
\citeauthor{2006ApJ...651..502P}, despite being fainter by some ten
magnitudes. The GOODS $v$-drop stars extend slightly blueward of the
bright sample in the [4.5]-[5.8] band, as would be expected given that
our $v$-drop sample extends to earlier M subclasses than the bright
sample (M3/M4 as opposed to M5). It is not clear from existing models
or data whether a similar bluewards shift might be expected with
varying metallicity.


While the bulk of the population has colours consistent with brighter
M stars, a subset of seven sources are anomalous with M star optical
colours and T star colours in the {\em Spitzer} wavebands. These fall
loosely into three categories.

One source, labelled `A' on figure \ref{fig:spitzer_14v23} has colours
consistent with a late-M/early-L classification within approximately
3\,$\sigma$ of the photometric errors in both optical and infrared
bands. The colours are also consistent with those expected for a close
M/L binary system with the M star flux dominating the optical bands
and the fainter L star contributing to the flux longwards of
3\,$\mu$m.  Binary M/L and M/T systems have been observed at bright
magnitudes in the past
\citep[e.g.][]{2002MNRAS.332...78L,2006AJ....131.1007B}. This source
is towards the faint end of our sample, with $i'_{AB}=24.12$. A binary
with separation of 1$''$.09 at a heliocentric distance of 35 pc
\citep[using the M/L binary pair of ][as a model]{2006AJ....131.1007B}
would be unresolved at {\em HST}/ACS resolution at the distance of
1\,kpc estimated for an M6 star at this magnitude.  If it is indeed a
binary system then each component will be intrinsically fainter still
and may lie at greater distances than estimated. Radial velocity
measurements of the system may be necessary to determine whether the
star is an anomalous M dwarf or a binary pair.

A further three sources, labelled `B' on figure
\ref{fig:spitzer_14v23} have $z'$-5.8\,$\mu$m colours consistent
with those of M stars but are anomalously bright in the 8\,$\mu$m
band. In one case, the photometric errors are large and the faint star
is within 2\,$\sigma$ of the colours expected for an M star.  However,
the other two stars are well detected in all bands, and show no
evidence of unreliable photometry. Both the 3.6 and 8\,$\mu$m bands
are dominated by CH$_4$ absorption with the fundamental band in the
bluewards filter. There is no clear interpretation of an 8\,$\mu$m
excess.

Finally the remaining three sources, labelled `C' on figure
\ref{fig:spitzer_14v23}, show no distinctive breaks or obvious
features in their spectral energy distributions but their flux peaks
further to the red than expected for M stars. Again these sources have
reliable photometry with no clear reason to believe photometric errors
are larger than shown.  As such they have M star colours in the
optical and T stars in the infrared.  Interestingly their $i'-[3.6]$
colours are well within the distribution of the rest of the sample,
showing no evidence for M/T multiplicity (which would lead to red
colours as the T dwarf contribution becomes dominant at 3.6\,$\mu$m).

We note that all of these anomalous sources defy straightforward
explanation based on photometry alone and may benefit from
spectroscopic analysis.

One unresolved object in the GOODS-N has extremely unusual colours in
the IRAC wavebands, lying outside of the region plotted in figures
\ref{fig:spitzer_23v12} and \ref{fig:spitzer_14v23}. The colours of
this source (($z'$-[3.6])$_{AB}$=0.71, ([3.6]-[4.5])$_{AB}$=0.09) may be
consistent with identification as a high redshift quasar or compact
galaxy and inconsistent with any cool star.

\section{The Properties of Cooler Stars in this Sample}
\label{sec:cooler}

To a limit of $z'_{AB}=26$, there are a total of 12 unresolved sources
in the GOODS fields that are identified as $i'$-drops but do not form
part of the $v$-drop sample due to their faint magnitudes in $i'$.

As figure \ref{fig:templates} indicated, an $i'$-drop selection is
sensitive to stars of spectral classes later than M8 (assuming
near-solar metallicities). The linear increase of $i'-z'$ colour with
spectral type becomes less clear cut at the same subclass, due to the
difficulty in assigning spectral types based on spectroscopic indices,
making classification of these sources impossible based on optical
photometry alone.

\begin{figure}
\includegraphics[height=0.9\columnwidth]{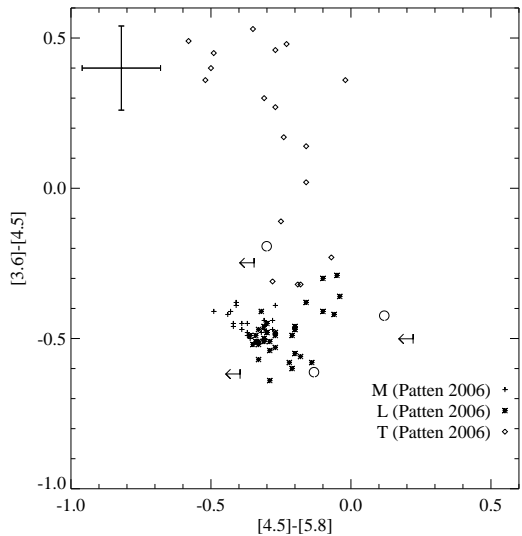}
\caption{The [4.5]-[5.8] versus [3.6]-[4.5] colours (open circles) or 3\,$\sigma$
  limits on the colours of isolated stars selected as $i'$-band
  dropouts. Given their $i'-z'$ colours, these sources are expected to
  be of class M8 or later.  Symbols as in figure
  \ref{fig:spitzer_23v12}.}\label{fig:idrops_23v12}
\end{figure}

Seven of these sources are unconfused in the 3.6\,$\mu$m band,
although given their faint magnitudes, one is undetected at
4.5\,$\mu$m a further four undetected longwards of 4.5\,$\mu$m.
Figure \ref{fig:idrops_23v12} shows the [4.5]-[5.8] versus [3.6]-[4.5]
colours or 3\,$\sigma$ limits on the colours of the six stars for
which these colours can be determined. All six are consistent with the
infrared colours of local M and L dwarf stars studied by
\citet{2006ApJ...651..502P}.

We note that an $i'$-drop selection used to identify cool stars
here is likely to be significantly incomplete for low metallicity dwarfs if the
available stellar models provide a reasonable approximation to their
colours, as figure \ref{fig:metallicity} illustrates.  A number of low
metallicity extreme subdwarfs have now been found. Unfortunately none
of these have published $z$-band photometry.

\citet{2007ApJ...657..494B} present the optical spectra and $I-J$
colours of a sample of late M and L extreme subdwarfs.  The
spectroscopically confirmed subdwarfs show $I-J$ colours between 1.4
and 1.8 for M dwarfs increasing to about 2.4 for L4 dwarfs. Given the
spectra of M class stars are essentially flat in f$_\lambda$ in the
range 8000-10000\AA, it seems unlikely that most late M subdwarfs
would have $i'-z'$ colours as extreme as 1.3. Indeed, synthetic
colours calculated from these spectra indicate that at these low
metallicities, late M dwarfs are approximately 0.2\,mag bluer in
$i'-z'$ than typical old-disk M stars (Bessell, private
communication). Nonetheless, the early L class extreme subdwarfs
of \citet{2007ApJ...657..494B} lie well redwards of the colour
selection criterion used here.

This metallicity-dependent selection effect complicates the
analysis of L and T star number counts, and larger surveys will be
needed to accurately constrain the surface density of these cool stars
at faint magnitudes..
Further progress in understanding the
incidence and properties of low metallicity class M, L and T dwarfs
may be possible in future given improved modelling of the effects of
varying metallicity on the spectrum longwards of 1\,$\mu$m and in the
{\em Spitzer}/IRAC wavebands.

\section{Spectroscopy of Distant M-dwarfs}
\label{sec:spect}

Some tens of faint low mass stars have now been observed in deep
spectroscopy, as part of the follow up to dropout-selected high
redshift surveys.

Extremely deep 8\,m class telescope spectra of M or L class stars with
$i'>22$ have been published or publically released by
\citet{2004ApJ...607..704S} and \citet{2005A&A...434...53V}.
Additional deep stellar spectra were obtained
as part of the ESO Remote Galaxy Survey
\citep[ERGS,][Douglas 2007, in prep]{2007astro.ph..1724D} and by
\citet{2003ApJ...593..630L} as part of their BDF survey. Many such
spectra remain unpublished, being considered incidental to the search
for $z>5$ galaxies.

Six of the $v$-drop sample discussed above have deep spectroscopy
obtained in multi-object mode with FORS2 on the VLT as part of the
GOODS programme and publically released by
\citet{2005A&A...434...53V}. These sources have magnitudes in the
range $i'_{AB}=23-24.5$ and were observed at a dispersion of
3.2\AA/pixel and a spectral resolution R=660, for a typical exposure
time of 14400 seconds. 

As discussed in section \ref{sec:photometry}, we supplement this small
sample with a further 18 faint M class star spectra observed as part
of the BDF project. The BDF survey comprises deep multicolour imaging
and spectroscopic follow-up of red sources in four near-contiguous
fields, applying an $R$-drop colour selection to identify galaxies at
$z>5$ and including a number of stars (which can be difficult to
separate from compact galaxies on the basis of optical colour alone).

$R$-drop sources were observed spectroscopically in multi-object mode.
Two FORS2 slitmasks were observed in each field, with a total integration
time of 13000 seconds per mask, divided into twenty exposures of 650\,s,
each offset by a different distance along the slitlets. Each slitlet
was 1 arcsecond wide. The spectra were wavelength calibrated from
arclamps and flux calibrated from standard star observations taken at
the same time. The final spectra span the spectral range
6000-10000\AA\ at a dispersion of 3.2\AA, and with a spectral
resolution R=660. This configuration is virtually identical to that
used by \citet{2005A&A...434...53V}, with the two samples differing
only in the selection filters.

\begin{figure}
\includegraphics[width=0.9\columnwidth]{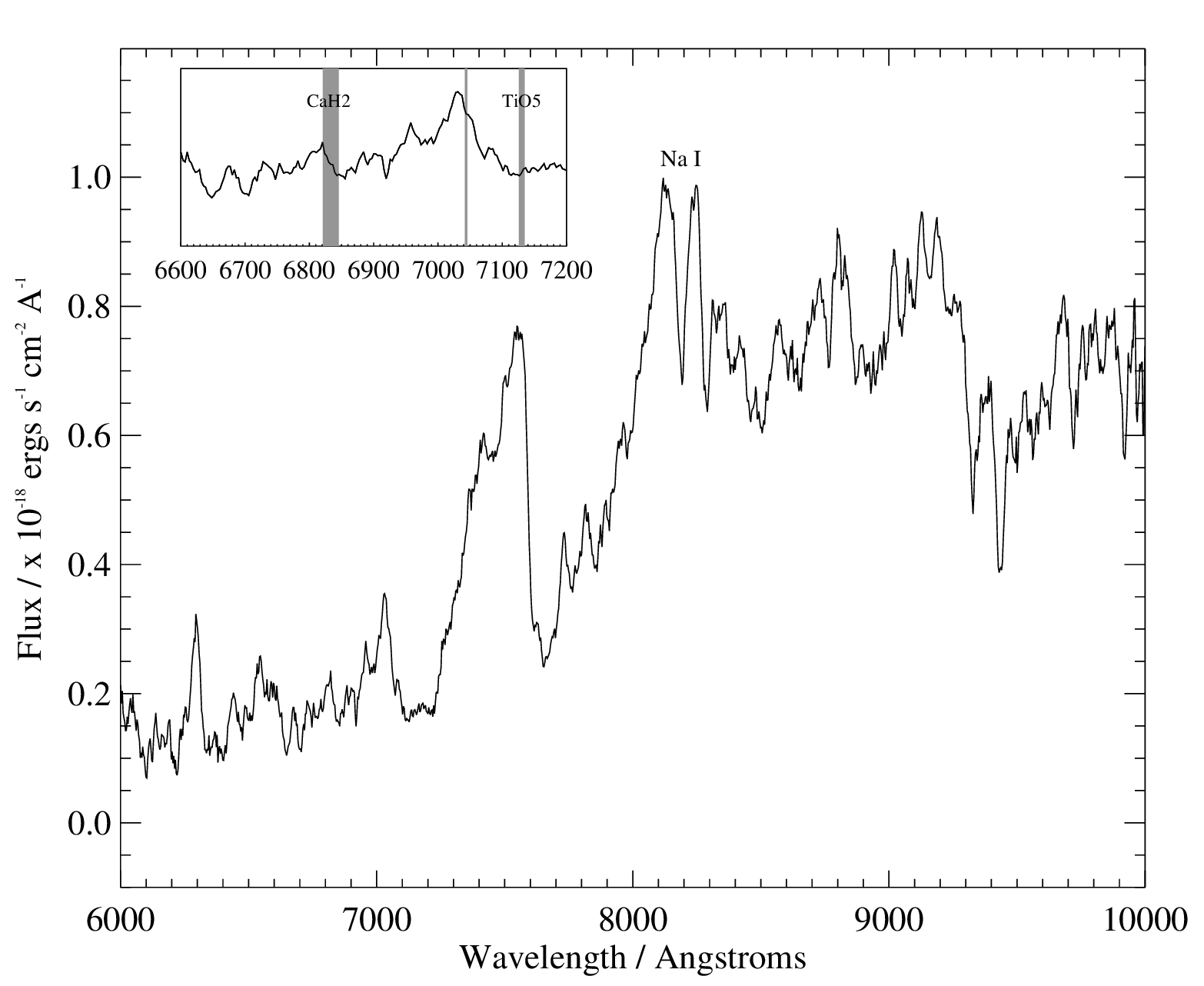}
\includegraphics[width=0.9\columnwidth]{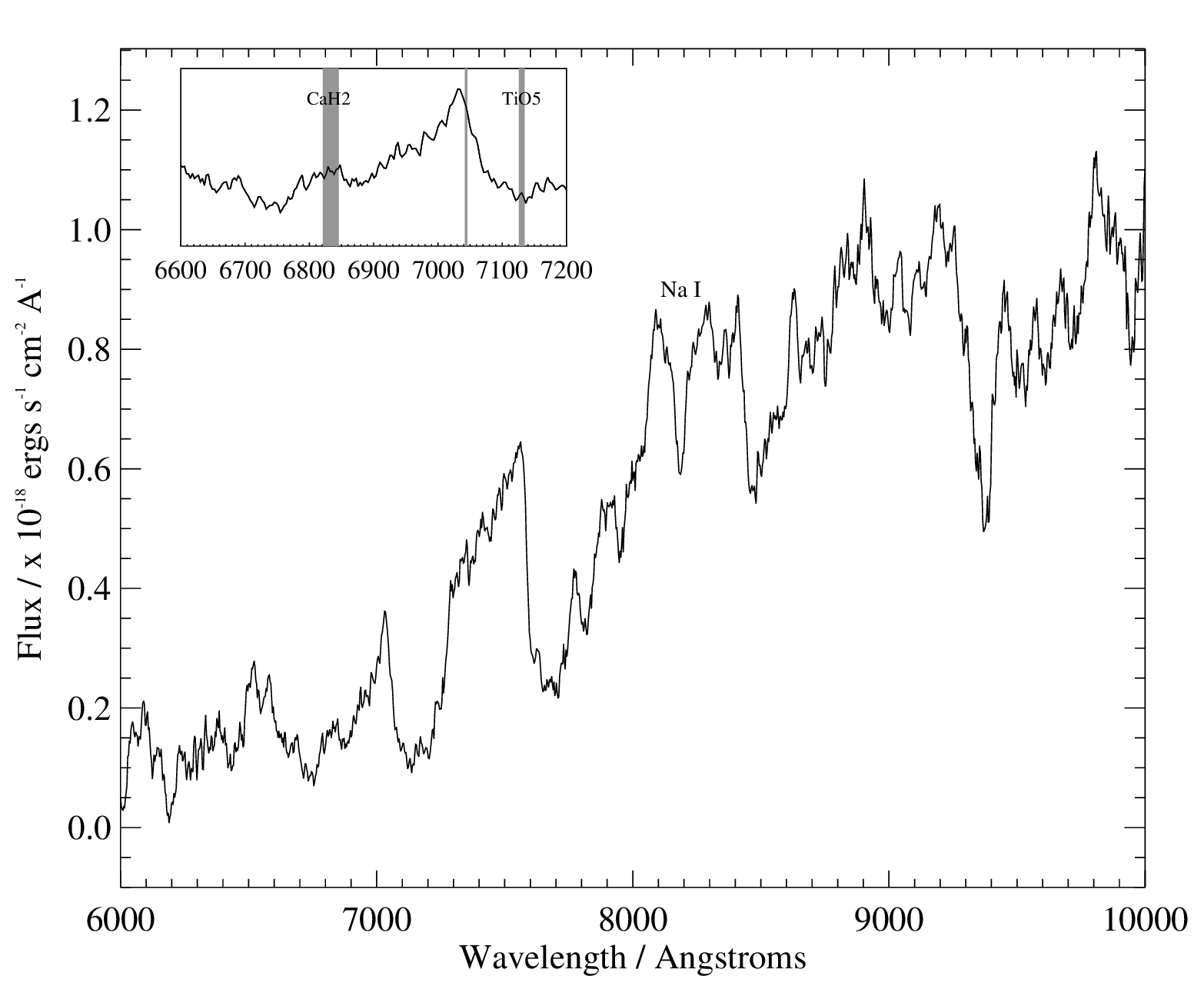}
\includegraphics[width=0.9\columnwidth]{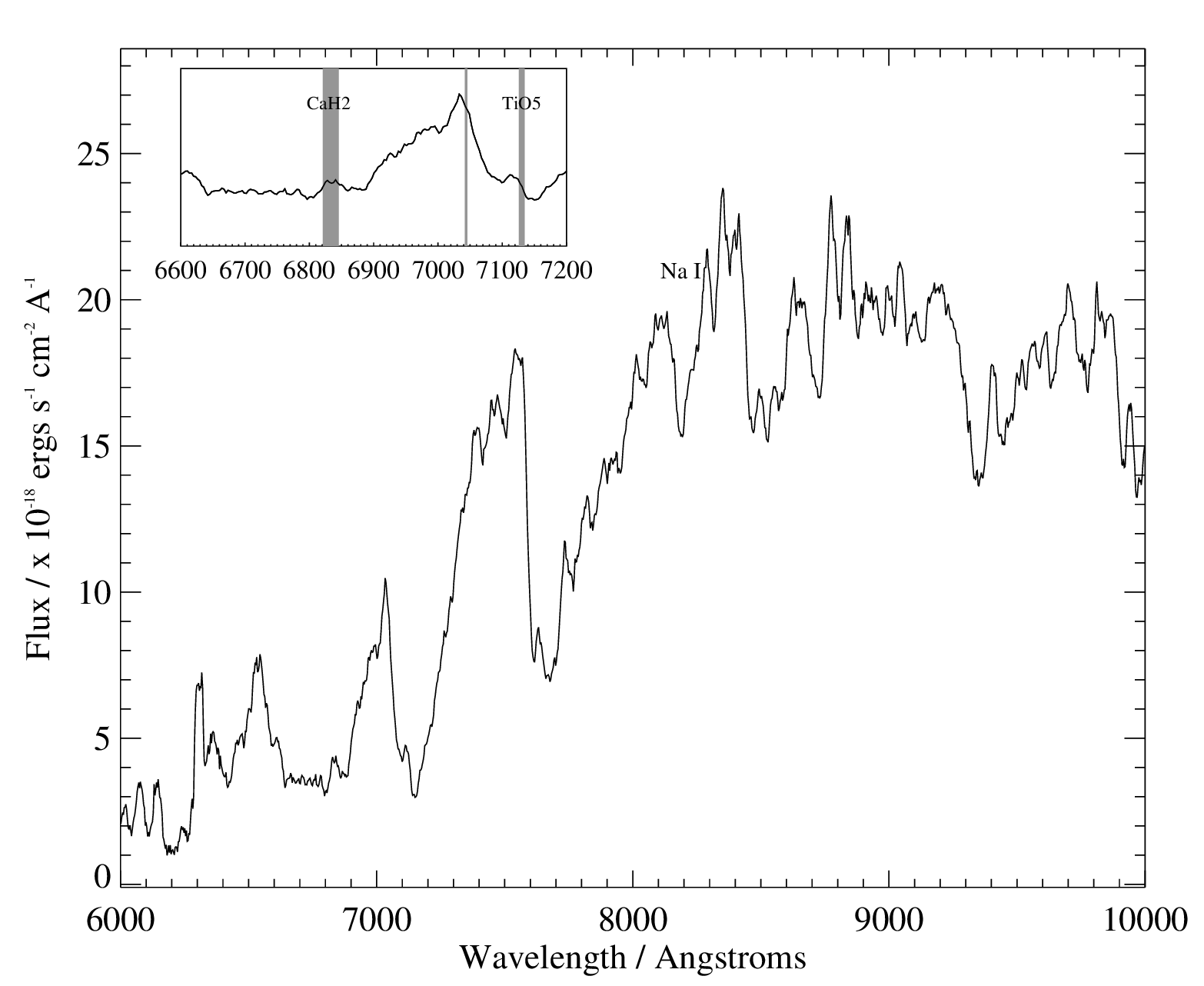}
\caption{Examples of stars identified in the BDF spectroscopy of
  an $R$-drop M star sample, as described in the text.  The absorption
  features of TiO 5, CaH 2 and Na I as defined in table
  \ref{tab:indices} are indicated by labels and/or grey shading on an
  inset magnification of the low wavelength
  features.}\label{fig:spec_examples}
\end{figure}

Examples of such faint star spectra are shown in figure
\ref{fig:spec_examples}. As expected for sources a hundred times
fainter than those observed by the SDSS, the spectra are noisy.
However, despite the low signal to noise of such spectra, they show
the similar molecular absorption features to their brighter
counterparts.  Each of the twenty-four stars in the final
spectroscopic sample discussed here has $i'_{AB}>23$ 
and signal to noise on the continuum around the
features of interest (i.e.  around 7000\AA) of 10 or
greater (in some cases higher than 20).

Given these deep spectra it is possible to identify a number of
molecular absorption features, both lines and absorption bandheads.
Here we consider two diagnostic features in the spectrum of M stars:
the surface gravity dependent Na I doublet at 8183, 8195\AA\ and the
metallicity dependent TiO to CaH ratio around 7000\AA.

The Na I doublet is a strong feature in the spectrum of M
stars, but its measurement is complicated by the crowded spectrum around
8200\AA\ which makes the true continuum level difficult to determine.
\citet{1980ApJ...235..405F} introduced a systematic measurement of the
equivalent width of this line by defining a region to contain the
line, and two adjacent regions as a proxy for the continuum level.
This definition is shown in table \ref{tab:indices}. 

As discussed by \citet{1997ApJ...479..902S}, this feature is a
sensitive indicator of surface gravity for stars cooler than 3900K,
and hence discriminates between dwarf stars with a surface gravity of
$\log(g)\approx5$ and giant stars of the same spectral class with a
typical surface gravity $\log(g)=1.0$. 

Giant stars of class M4 and later are rare, and not expected to
substantially affect the number counts of faint stars, but the presence
of a significant subpopulation of giants could theoretically impact
our distance distributions (which are based on distance modulus,
assuming that the stars are dwarfs).  As expected, the Na I equivalent
width of all eighteen spectra are entirely consistent with dwarf stars
(having EW$>5$), and inconsistent with those of M giants (which would
have an EW$<1.5$).

\label{sec:indices}

Given their optical colours (discussed in section \ref{sec:halo-pop})
the M dwarfs described here are expected to be comparable to or
slightly lower in metallicity than M dwarfs observed in a local
sample.  To test this hypothesis, we consider the ratio of two
molecular line indices. Low metallicity dwarfs have lower ratios of
CaH to TiO than higher metallicity dwarfs \citep{1997AJ....113..806G},
allowing their ratio to be used as a crude metallicity indicator.

The strength of these lines is usually gauged by a series of
narrowband indices first measured by \citet{1995AJ....110.1838R}.
These indices measure the strength of an absorption feature relative
to a pseudo-continuum flux measured nearby. The definition of the TiO
5 and CaH 2 indices used here is given in table \ref{tab:indices}.

\begin{table}
\begin{tabular}{lcc}
Band & S1 & W\\
\hline\hline
TiO 5 & 7042 - 7046\AA & 7126 - 7135\AA \\
CaH 2 & 7042 - 7046\AA & 6814 - 6846\AA \\
\\
Na I EW & 8169-8171\AA & 8172-8209\AA \\
        & 8209-8211\AA & \\
\end{tabular}
\caption{The regions used in spectroscopic indices defined by
 \citet{1995AJ....110.1838R} and the Na I doublet equivalent
  width defined by \citet{1980ApJ...235..405F}.
 Indices are defined as $R=F_W/F_{S1}$ where the spectral region
 $W$ measures the line strength and the sidebar $S1$ is indicative
 of the continuum strength.}\label{tab:indices}
\end{table}

\begin{figure}
\includegraphics[width=0.9\columnwidth]{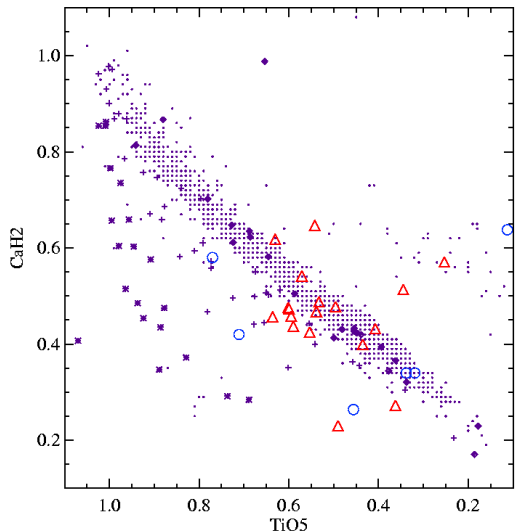}
\caption{The spectroscopic abundance indices of faint M-stars identified
  during spectroscopic follow-up of high redshift dropout samples.
  Blue circles indicate the GOODS sample of
  \citet{2005A&A...434...53V} while red triangles show the BDF sample
  presented in this work. Typical errors on these indices are better
  than 10\%. The spectroscopic indexes of local M and L stars are shown for
  reference. Points indicate the local (solar metallicity) sample of
  \citet{1996AJ....112.2799H} and \citet{1995AJ....110.1838R} while
  small symbols indicate the sample of \citet{1997AJ....113..806G}
  (diamonds -- solar metallicity, crosses -- [Fe/H]=-1, asterisks -
  [Fe/H]=-2).}\label{fig:indices}
\end{figure}

On figure \ref{fig:indices} we plot the ratio of line indices measured
on the 18 spectra of the combined GOODS and BDF samples. There is
little evidence for significantly sub-solar metallicities (i.e.
[Fe/H]$\approx$-2). Around 40\% of the sample (7 of 18) are consistent
with slightly sub-solar metallicities (i.e. [Fe/H]$\approx$-1), while
the same fraction are completely consistent with the indices for a
local (solar metallicity) sample. 20\% of the sample (4 stars) show
anomalously strong CaH features given the strength of their TiO
absorption bands. Of these, two also show TiO5$<0.3$ suggesting that
they are of spectral type M6 or later (consistent with their
photometric colours).

We caution the reader that it is possible that these spectral indices
may be affected by noise due to the subtraction residuals of night sky
emission, and by night sky absorption, which isn't removed by our data
reduction. Spectroscopy aimed at identifying high redshift galaxies is
routinely obtained in such a manner as to optimise night sky line
subtraction. By offsetting the telescope between exposures, such that
the target moves along the slit, the frames can be combined during
reduction in such a way as to produce a simultaneous measurement of
the sky flux that can be subtracted from the object spectrum. The
measured signal to noise on the spectra account for the error in the
sky line subtraction, and hence given our requirement of a signal to
noise exceeding ten, we estimate statistical error on our spectral
indices of less than 15\%. However, we note that the indices are
measured in a spectral region affected by a weak skyline complex, and
that our spectra are of comparatively low resolution.

While we do not attempt further quantitative analysis of our spectra,
we note that there are other spectral features that may be accessible
even in these faint sources. In particularly, it may be possible in
future to measure the metallicity using the KI doublet feature at
7700\AA. This doublet is observed to be extremely strong in the low
metallicity extreme subdwarfs of \citet{2007ApJ...657..494B}. While
the line is not as strong in early M dwarfs as in later subtypes, we
note that our spectra appear qualitatively similar to those from the
nearby near-solar metallicity with no sign of the enhancement in line
strength that might be expected at low metallicities.

The typical metallicity of the sample as derived from 18 spectra is
consistent with that determined from their optical colours (figure
\ref{fig:metallicity}b).  Given that none of the eighteen stars in our
spectroscopic sample are of very low metallicity ([Fe/H]$<$-2.), the
fraction of significantly sub-solar sources in our photometric sample
is likely to be small ($<6$\%), consistent with the suggestion of two
sources out of 124 ($<2$\%) based on their optical colours.

\section{A Metal-rich Population of M-class Stars at Halo Distances}
\label{sec:phot-properties}

Given the evidence of their optical and infrared colours, combined
with photometry of a number of examples, it appears that the M star
population identified at faint magnitudes (and hence in extragalactic
surveys) is relatively metal-rich ([Fe/H]$>$-1) compared to the
metal-poor halo population that might be expected to dominate star
counts at large distance. In fact, their metallicities are more
typical of those observed locally in the thick disk.  This suggests
that there is only a weak metallicity gradient in the class M and
later dwarfs identified as photometric dropout sources, although
this analysis does not exclude the possibility that a low
metallicity M-star population also exists at faint magnitudes.

Given this relatively metal-rich population, it is possible to explore
the comparison with brighter M stars further.  As demonstrated by
\citet{2002AJ....123.3409H}, for dwarf stars with metallicities
[Fe/H]$>$-1, $i'-z'$ colour increases linearly with subclass between
spectral classes M3 and M9, flattens through the early L dwarfs and
then increases again.  As a result, M stars identified as $v$-drops
(which are typically too blue in $i'-z'$ to be L stars) can be crudely
classified by their $i'-z'$ colours. We chose to use $i'-z'$ colour
for this classification rather than $v-i'$ for two reasons: in order
to minimise errors on the colour, and in order to provide a second
colour, thus beginning to constrain the SED for these sources.  In the
deep optical imaging employed here, the typical photometric errors on
this colour are of order 0.05 mag, less than the measured variation of
colours within each subclass in a brighter sample from the SDSS
(determined by comparing spectral typing based on template fitting,
spectral indices and photometric colour) which is typically 0.2 mag in
these filters,\citep{2007AJ....133..531B,2002AJ....123.3409H}. This
scatter in $i'-z'$ colour within subclasses arises from variation in
physical properties of the stars in question. A continuum in the
metallicity and temperature of individual stars will affect their line
indices and photometry, leading to a gradual range of characteristics
in a field population rather than the discrete set of subtypes that
might be expected from a single age, single metallicity population
such as a globular cluster. This uncertainty in intrinsic colour
limits photometric classification to $\pm1$ subclass. The surface
density of faint stars decreases rapidly with increasing M star
subclass, since a smaller volume is probed to the same magnitude
limit. As a result stars in the range M4-M5 are expected to outnumber
later classes.

\begin{figure*}
\includegraphics[width=0.9\columnwidth]{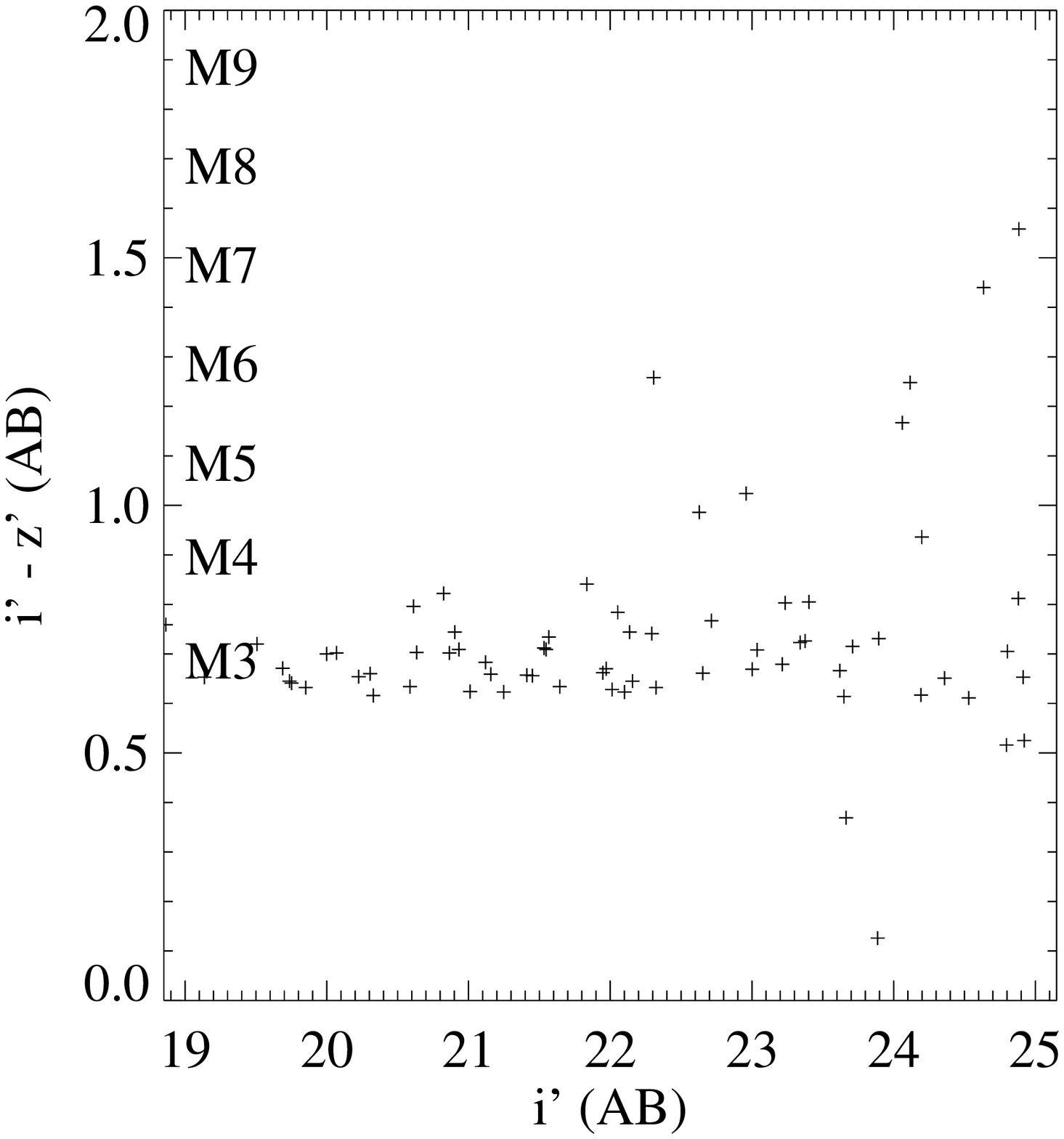}
\includegraphics[width=0.9\columnwidth]{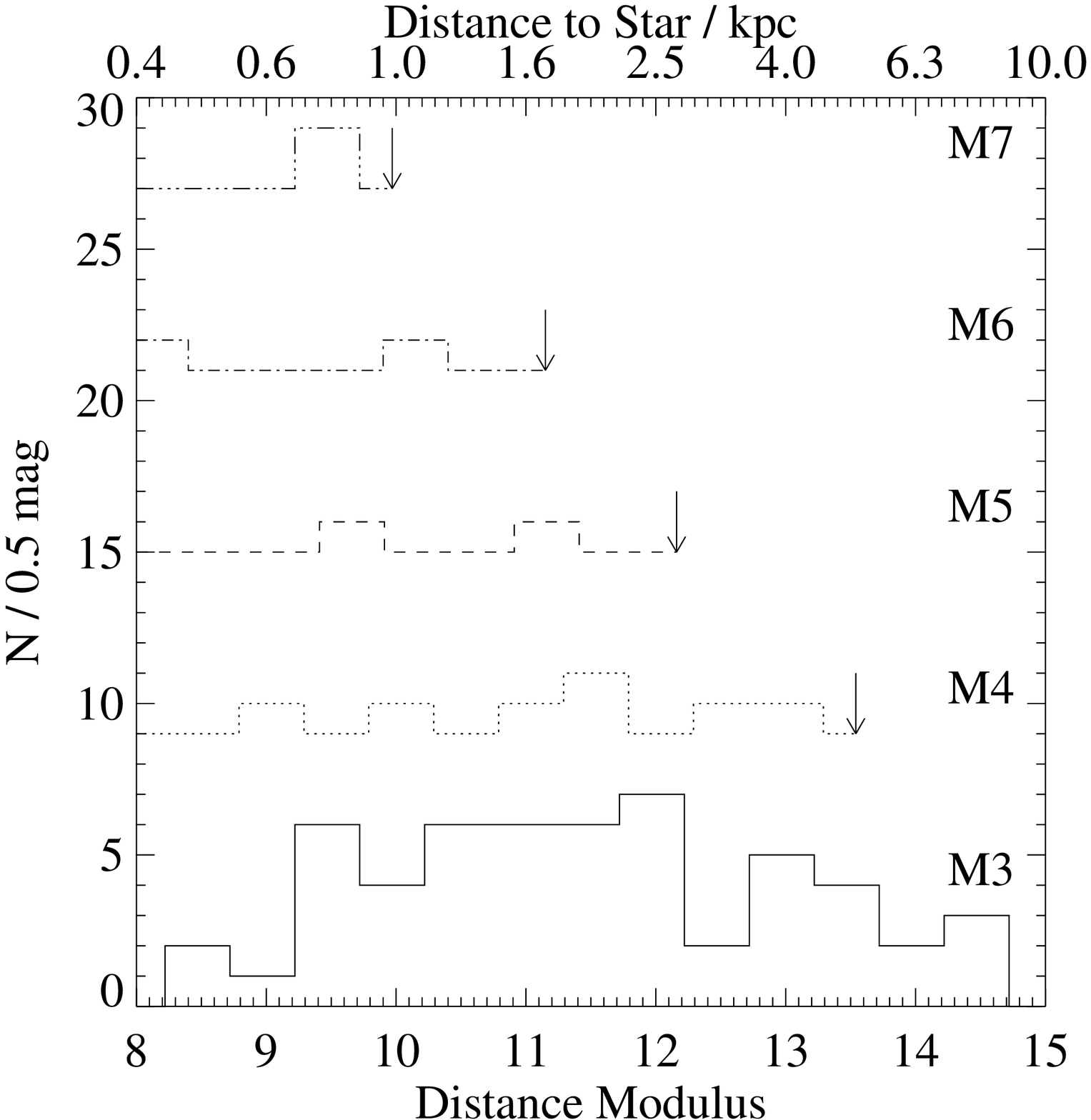}
\caption{(Left) The $i'-z'$ colours of $v$-drop selected low
  mass stars in the GOODS-N field. M-star subclass increases linearly
  with $i'-z'$ colour \citep{2002AJ....123.3409H,2005PASP..117..706W}
  enabling a crude classification of stars based on their photometry.
  (Right) The distance to those stars, classified into subclasses by
  $i'-z'$ colour and applying the spectroscopic parallax calibrated
  typical absolute magnitudes for each subclass as determined by
  \citet{2005PASP..117..706W}. The maximum distance probed by our
  photometric sample for each subclass is indicated by an
  arrow.}\label{fig:data_n}
\end{figure*}

\begin{figure*}
\includegraphics[width=0.9\columnwidth]{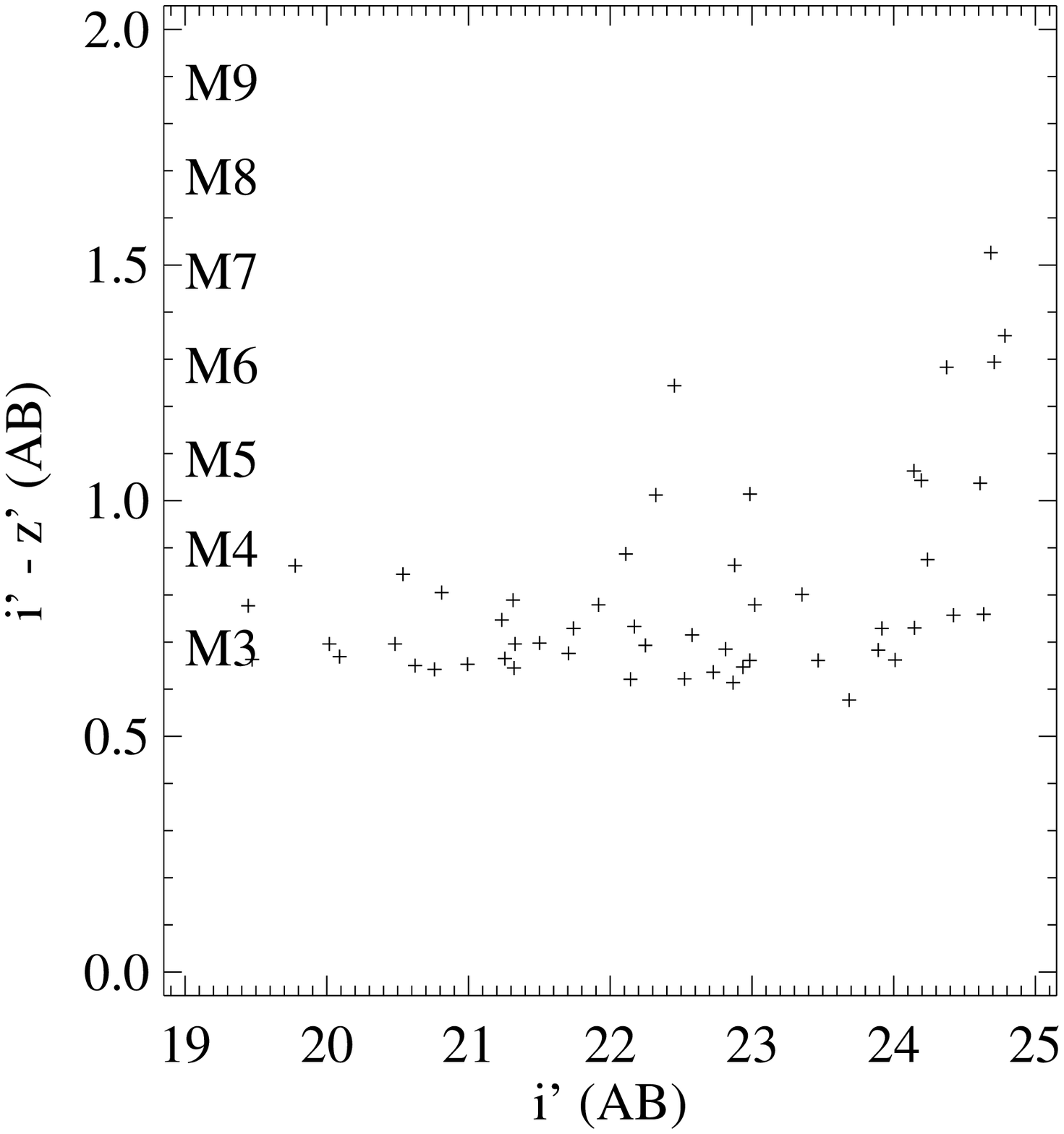}
\includegraphics[width=0.9\columnwidth]{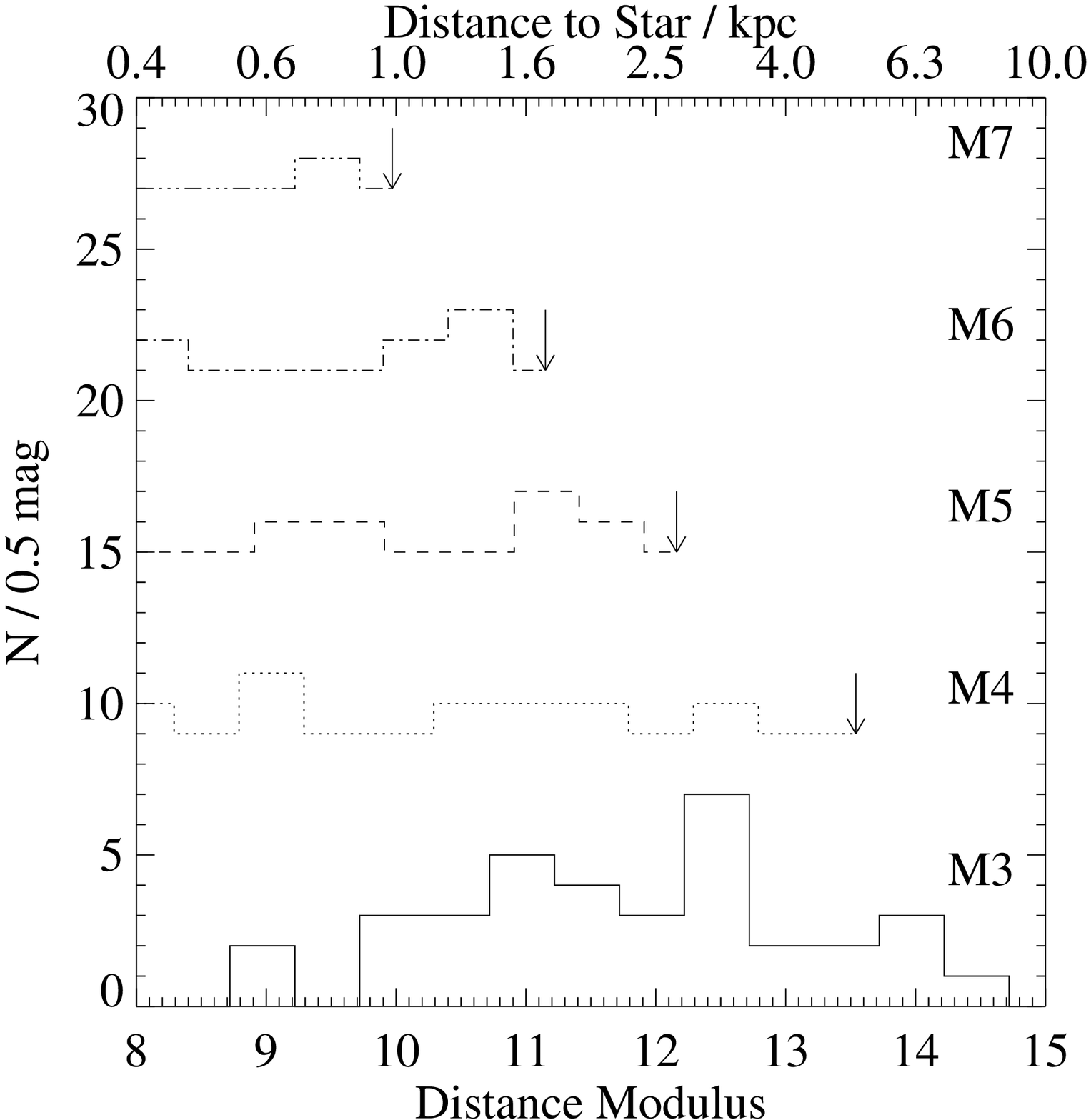}
\caption{(Left) The $i'-z'$ colours of $v$-drop selected low
  mass stars in the GOODS-S field. (Right) The distance to those
  stars, classified into subclasses by $i'-z'$ colour and applying the
  typical absolute magnitudes for
  each subclass as determined by \citet{2005PASP..117..706W}. As in
  figure \ref{fig:data_n}.}\label{fig:data_s}
\end{figure*}

We make an approximate classification of our sample of $v$-drop M stars
by assigning each to the closest subclass in typical colour, based on
the templates of \citet{2007AJ....133..531B}.  As figures
\ref{fig:data_n} and \ref{fig:data_s} illustrate, early M stars
(classes M3-M4) are found spanning the entire range from the
saturation limit to the faintest magnitudes at which the GOODS data
can distinguish unresolved sources from compact galaxies.

Later M stars (spectral class M5 and later) are intrinsically fainter
and so our survey probes progressively smaller volumes with increasing
subclass. These are not seen at bright magnitudes but become
increasingly common at $i'>22$. The latest stellar class seen in our
$v$-drop samples is M7. Stars of classes later than M7 are
intrinsically faint and so will only be detected in a relatively small
volume probed locally. In addition, the very red $i'-z'$ colours of
these sources is likely to lead to them dropping below the limit of
the $i'$ band used for selection. Later classes may be detected rather
as $i'$-band dropouts.

We note that M3 stars should formally be too blue to be selected given
our $v-i'$ colour cut.  As discussed above, classification by $i'-z'$
colour is at best an approximate method, and the large `M3' sample may
be explained as a combination of M4 stars with bluer than average
$i'-z'$ colour and genuine M3 stars with a redder than average $v-i'$
colour. Spectral classification would be necessary to classify
individual sources. Interestingly these `M3' sources are more numerous
than the `M4' sample suggesting that scatter in the $i'-z'$ colour is
skewed to the blue rather than symmetrical which may be indicative of
low metallicity as discussed in section \ref{sec:halo-pop}. If so, a
slight correction must be applied to the absolute magnitude of each
subclass resulting in slightly lower estimated distances. However,
while this will apply a systematic shift to the distribution of
distance moduli, it will not change the distribution.

As expected, the two probable low metallicity candidate $v$-drops in
the GOODS-N field defy classification based on $i'-z'$ colour.

Using the typical absolute magnitude for each class \citep[as
determined by ][]{2007AJ....133..531B} we can estimate the distance
modulus, and hence distance to each star, essentially using the
relation between colour and absolute magnitude determined from the
local disk population \citep{2007AJ....133..531B,2002AJ....123.3409H}
to derive a photometric parallax for each star in the sample. While
this has traditionally been carried out using the $V$ vs $V-I$
relation, we chose instead to use $i'$ vs $i'-z'$ \citep[in common
with][]{2002AJ....123.3409H} since these sources are brighter, and
hence their magnitudes are more reliable, in the redwards bands.

The distribution of source
distances from the sun are shown in the right-hand panels of figures
\ref{fig:data_n} and \ref{fig:data_s}. We are observing class M3-M4 stars
at distances between 1 and 14\,kpc. The distant tail of this population
is well above both the scale height of 275\,pc for the thin disk, and the 
1.5\,kpc scale height of the thick disk that may be more appropriate for
disc M dwarfs \citep{2001ApJ...555..393Z}.

The most extensive survey of faint M dwarfs to date has been a survey
of high galactic latitude fields with {\em HST}/WFPC
\citep{1994ApJ...435L..51B,2001ApJ...555..393Z}. These authors probed
about 1400 stars, reaching a depth of $I_\mathrm{Vega}=23.7$
($I_{AB}=23.2$). Their finding of a sharp decline in the number of
stars with a relatively short scale height of $<$ 1\,kpc above the
galactic disc \citet{2001ApJ...555..393Z} seems inconsistent with our
flat number counts out to a heliocentric distance of $10$\,kpc.
Clearly our sample is smaller by a considerable margin, but is also
significantly deeper and hence highly complete ($>$98\%) at large
heliocentric distances. The fact that a similar distribution of
number counts is observed in two widely separated fields argues against
Galactic substructure as an explanation.

A flat number density of stars with increasing magnitude is not easily
reproduced given standard models of the galactic density profile. In
order to explore this, we constructed Monte Carlo simulations of a
sample of 50 stars randomly drawn from a population of $1 \times 10^6$
stars distributed according to a known density profile.  Profiles
considered include the thick disc density law of
\cite{2001ApJ...555..393Z} which takes the form of a sech$^2$(z/h) law
with a scale height h=300\,pc above the Galactic disk and `model B' of
\cite{1996AJ....112.1472R} which considers a composite model of thin
disk and thick disc (both with sech$^2$ density profiles and scale
heights of 350pc and 1.5kpc respectively) together with a halo density
profile (scaling as approximately r$^{-3.5}$ with radius from the
galactic centre). Variations on these models were also considered with
varied scaling between disc and halo components, and a variety of halo
power laws. In each case, number counts are predicted to increase
sharply at faint magnitudes, since the density law varies less rapidly
than the volume element at a given magnitude. The absence of such an
upturn suggests that this population is falling off more rapidly than
those expected from standard disk profiles, despite extending many
scale heights above the plane of the disk, or alternately that the
conventional components of galactic structure need to be rethought.
Our results might be interpreted as suggest that either the scale
height of the thick disk has been significantly underestimated (as
also suggested by \citet{1996AJ....112.1472R}) or that a third distant
but relatively metal-rich component is needed to explain the
population. However, we do caution that our stellar number counts in
the most distant bins are small and that analysis of larger deep
surveys are needed. 

We note with interest that the presence of a large population of
massive astrophysical compact halo objects (MACHOs) that are faint in
the optical and contribute up to 20\% of the halo mass has been
inferred from microlensing analyses \citep[e.g.][and references
therein]{2007A&A...469..387T,2005ApJ...633..906B,2000ApJ...542..281A}.
These sources have been inferred to have a mass of
$\sim$0.5\,M$_\odot$, similar to the mass of an early to mid M-dwarf
star.  According to the widely used `S'-model
\citep{2000ApJ...542..281A}, the MACHO objects occupy a spherical,
isothermal halo of core radius 5\,kpc. In at least one case, a lensing
star has been confirmed photometrically and spectroscopically to be an
M4-5 dwarf star, with a red colour similar to those discussed in this
analysis \citep[$V_{F555W}-I_{F814W}=3.2$
(Vega),][]{2001Natur.414..617A}.  While the M dwarf population studied
here may explain only a fraction of the proposed MACHO mass
contribution to the dark halo, we probe only a small subset of the
halo M star population with the colour and flux limits applied in our
study.  We note that these cool, red stars would not easily be
identified in the relatively shallow and blue monitoring imaging
employed by the MACHO \citep{2000ApJ...542..281A} and EROS
\citep{2007A&A...469..387T} collaborations.


Our survey also highlights both the utility and the difficulty of
relying on the relatively small deep fields designed for extragalactic
observations. While such surveys are extremely sensitive and benefit
from multiwavelength observations - often including coordinated
observations in the infrared and deep spectroscopic follow-up - they
also suffer from being necessarily small in size. The newer generation
of ground-based, wide-field imagers have enabled larger deep fields
to be observed, but under seeing conditions that make the effective
separation of stars and high redshift galaxies (which become
comparable in number density at faint magnitudes) impossible. The
variation of 40\% in the M star number counts between the GOODS fields
suggest that larger deep fields should be observed at {\em HST} resolutions
in order to average out structure in the inner halo or outer disc.

Clearly further work in larger fields with both deep and high
resolution imaging \citep[e.g. COSMOS,][]{2006astro.ph.12305S} is
required to strengthen the observational result that the number
density of sources remains constant to faint magnitudes, which may
imply a rapid cut-off in the thick disk density profile at large
distances above the Galactic plane.

\section{Conclusions}
\label{sec:conclusions}

We have investigated the properties of a faint population of M stars
selected using a photometric drop method similar to that used to
identify high redshift galaxies.  The M-dwarfs in this study are up to
five magnitudes fainter than those considered by previous authors, and
hence lie at distances up to ten times further than those investigated
before.

This population extends well above the galactic plane, with little
evidence for a decline in number at faint magnitudes/large distances.
This is difficult to reconcile with models of known disk structure.

The M-stars selected by this photometric method show similar colours
to those within a kiloparsec of the sun, both in the optical and the
infrared.  Together with the evidence from line indices, this
suggests that there are significant numbers of M-stars with moderate
to high metallicities extending well above the plane of the galaxy.

Our sample of stars with $v-i' > 1.7$ should be complete for mid-M
dwarfs of near-solar metallicity but is incomplete for low-metallicity
M dwarfs many of which have bluer colours or are more distant.
Determining the metallicity of individual sources is difficult without
exceptionally deep spectroscopy. If one or more of our sources are at
low metallicity, they may be even cooler - and hence somewhat closer -
than we believe, although our spectroscopic results indicate that
fewer than 6\% of M stars are likely to fall into this category.


\section*{Acknowledgements}

ERS gratefully acknowledges support from the Particle Physics and
Astronomy Research Council (PPARC). The authors thanks Gerry Gilmore,
Laura Douglas and Avon Huxor for useful discussions. We also thank our
referee, Mike Bessell for comments and suggestions that have improved the paper.

Based on observations made with the NASA/ESA Hubble Space Telescope,
obtained from the Data Archive at the Space Telescope Science
Institute, which is operated by the Association of Universities for
Research in Astronomy, Inc., under NASA contract NAS 5-26555. This
work is also based in part on observations made with the Spitzer Space
Telescope, which is operated by the Jet Propulsion Laboratory,
California Institute of Technology under a contract with NASA. We
thank the GOODS team for their hard work in making these high quality
datasets public.

Results from the BDF fields was based on observations made with ESO
telescopes at the Paranal Observatory under programme IDs 69.A-0656
and 71.A-0290.

\label{lastpage}


\end{document}